\documentclass[]{aastex631}
\usepackage{newtxtext,newtxmath}

\usepackage{lastpage}
\usepackage{booktabs}
\usepackage{longtable} 
\usepackage{lipsum} 
\usepackage{array} 

\usepackage[T1]{fontenc}
\usepackage{CJKutf8} 
\begin{document}

\title{The redshift evolution of the luminosity function of type II GRBs}

\author[0000-0002-9838-4166]{Yan-Kun Qu}

\affiliation{School of Physics and Physical Engineering, Qufu Normal University, Qufu, 273165, People's Republic of China}
\author{Zhong-Xiao Man}
\affiliation{School of Physics and Physical Engineering, Qufu Normal University, Qufu, 273165, People's Republic of China}
\author[0000-0002-0786-7307]{Yu-Peng Yang}
\affiliation{School of Physics and Physical Engineering, Qufu Normal University, Qufu, 273165, People's Republic of China}
\author[0000-0003-0672-5646]{Shuang-Xi Yi}\footnote{yisx2015@qfnu.edu.cn(YSX)}\footnote{quyk@qfnu.edu.cn(QYK)}
\affiliation{School of Physics and Physical Engineering, Qufu Normal University, Qufu, 273165, People's Republic of China}
\author{Mei Du}
\affiliation{School of Physics and Physical Engineering, Qufu Normal University, Qufu, 273165, People's Republic of China}
\author{Fa-yin Wang}
\affiliation{School of Astronomy and Space Science, Nanjing University, Nanjing, 210093, People’s Republic of China}
\affiliation{Key Laboratory of Modern Astronomy and Astrophysics, Nanjing University, Nanjing, 210093, People’s Republic of China}



\begin{abstract}
As of December 2023, the Swift satellite has detected more than 1600 gamma-ray bursts (GRBs). 
 We select 307 Type II GRBs for constructing the luminosity function (LF) based on the following criteria: (1)  duration $T_{90} \geq 2 s$; (2) conformity with the Amati relation for Type II GRBs; and (3)  peak flux $P \geq 1 \, \text{ph} \, \text{cm}^{-2} \, \text{s}^{-1}$. We explore two general forms of the GRB LF: a broken power-law (BPL)  LF and a triple power-law (TPL)  LF. We consider three evolutionary scenarios: no evolution, luminosity evolution, and density evolution. We find that the no evolution model can be excluded, while both luminosity and density evolution models effectively account for the observations. This result is consistent with previous studies on long GRBs (LGRBs). However, our Type II GRB sample favors a BPL LF, in contrast to the preference for a TPL function discovered in LGRBs.

\end{abstract}



\section{Introduction} \label{sec:intro}
Gamma-ray bursts (GRBs) are among the most energetic explosive events in the universe, detectable across vast cosmic distances, from nearby regions to the farthest observable areas \citep{2009Natur.461.1258S,2013FrPhy...8..661G,2017AdAst2017E...5C}. 
The highest recorded redshift for GRBs is $z\sim 9.4$ \citep{2011ApJ7367C}, and with the successful launch of SVOM in June 2024, it is now expected that GRBs will be detected at redshifts as high as $z\sim 12$ \citep{2024arXiv240303266L}, highlighting their significant potential as probes for cosmological studies. 
Since the launch of the \textit{Swift} satellite in 2004 \citep{2004ApJ...611.1005G}, more than 1600 GRBs have been detected, with redshift measurements available for about 500.

GRBs are generally categorized by their duration \(T_{90}\): LGRBs, lasting more than 2 seconds, and short GRBs (SGRBs), lasting less than 2 seconds \citep{1981Ap&SS..80....3M,1993ApJ...413L.101K,2013ApJ...764..179B}. LGRBs are believed to originate from the collapse of massive stars, supported by the association of some LGRBs with supernovae \citep{1998Natur.395..670G,2008ApJ...687.1201K,2010MNRAS.405...57S}, and their frequent occurrence in star-forming regions \citep{1997ApJ...486L..71T,1998ApJ...507L..25B}. In contrast, SGRBs are thought to result from mergers of compact binary systems, with evidence including their association with kilonovae \citep{2013Natur.500..547T,2019MNRAS.489.2104T,2020NatAs...4...77J} and gravitational waves \citep{2017ApJ...848L..13A}.

If the standard models of GRBs hold true, they could serve as reliable standard candles at high redshifts. Studies using GRBs to constrain cosmological parameters often rely on empirical correlations observed within GRBs themselves \citep{2015NewAR..67....1W}. However, the precision of these constraints is typically an order of magnitude lower than that achieved with Type Ia supernovae \citep{1998AJ....116.1009R}.  Although the efforts have been made to improve GRB-based cosmological constraints by focusing on GRBs with similar spectral characteristics, such as those exhibiting a plateau phase \citep{2022ApJ...924...97W,2023ApJ...953...58L,2024A&A...689A.165L,2023ApJ...958...74T}, these approaches have yet to reach precision comparable to that of Type Ia supernovae.

A fundamental issue is the relationship between the GRB event rate and the star formation rate (SFR). LGRBs are widely believed to result from the supernova explosions of massive Wolf-Rayet stars \citep{2010NewAR..54..206L}, with their occurrence rate expected to closely mirror the SFR. However, this correlation remains debated. Some studies report a consistent alignment between the GRB occurrence rate and the SFR \citep{2012A&A...539A.113E,2013ApJ...772...42H,2016A&A...587A..40P}. Conversely, other studies suggest that the GRB event rate significantly exceeds expectations, particularly at high redshifts \citep{2008ApJ...673L.119K,2008ApJ...683L...5Y,2009MNRAS.400L..10W,2011MNRAS.417.3025V}, while others indicate an excess at lower redshifts \citep{2015ApJ...806...44P,2015ApJS..218...13Y,2017ApJ...850..161T,Zhang2018,2019MNRAS.488.5823L,2022MNRAS.513.1078D}.

\cite{2021MNRAS.508...52L} found that when constructing the LF of Swift LGRBs, the TPL model provides a significantly better fit compared to the BPL model, which is also found by \cite{2015ApJ...812...33S}. BPL or single power-law (SPL)  are more common  forms of LFs \citep{2015ApJS..218...13Y,2015ApJ...806...44P,2019MNRAS.488.5823L}, while the TPL is relatively rare. \cite{2021ApJ...914L..40D} suggests that the TPL LF could possibly be caused by an excess of low-redshift sources. \cite{2024ApJ...963L..12P} found that at low redshifts (z < 2), the event rate of LGRBs exceeds the star formation rate (SFR), and the excess resembles the shape of a delayed SFR. In contrast, at high redshifts, the event rate of LGRBs aligns more closely with the SFR. \cite{2024ApJ...976..170Q} hypothesized that all high-redshift GRBs originate from the collapse of massive stars and fit the LF of high-redshift GRBs. When this LF is extended to low redshifts, it significantly underestimate the observed number of low-redshift LGRBs.  This strongly supports the idea that a considerable fraction of low-redshift LGRBs may not originate from massive star collapses.
The LF of collapsar GRBs is generally well described by a BPL model.
In contrast, \cite{2015ApJ...812...33S} found that the LF of merger-origin GRBs can be described by a SPL model. It is well known that merger-origin GRBs tend to have lower average luminosities compared to collapsar GRBs \citep{2014ARA&A..52...43B}. If LGRBs are a combination of higher-luminosity collapsar GRBs that follow a BPL LF and lower-luminosity merger-origin GRBs that follow an SPL LF, it naturally leads to the formation of a TPL LF.  In this study, we select a pure sample of Type II GRBs to rigorously test the LF models.

When analyzing the LF of GRBs using Swift data, two significant challenges must be addressed. The first challenge pertains to selection effects. Observational samples are inherently subject to various instrumental biases and selection effects, which complicate the task of accurately uncovering the intrinsic distribution and evolutionary characteristics of LGRBs \citep{2003AJ....125.2865B}. In particular, not all faint GRBs that barely surpass Swift's detection threshold will successfully trigger the detector, further skewing the observed sample. The second challenge involves redshift completeness. While the redshift completeness of Swift-detected GRBs is approximately 30\% \citep{2013ApJ...764..179B}, a substantial improvement over earlier detectors like BATSE \citep{1993ApJ...413L.101K} and Konus-Wind \citep{1981Ap&SS..80....3M}, the observed redshift distribution may still not faithfully reflect the true intrinsic distribution of GRBs.

To tackle these challenges, two distinct approaches can be employed. The first approach leverages a sample characterized by high redshift completeness and high luminosity. \cite{2012ApJ...749...68S} proposed a set of criteria for selecting a complete high-redshift sample. Before May 2011, of the GRBs detected by Swift, 58 met these criteria, with 52 having associated redshifts, resulting in a redshift completeness of 90\%. \cite{2016A&A...587A..40P} refined this sample, identifying 99 GRBs, of which 82 had redshift measurements, yielding a redshift completeness of 82\%. Despite the robustness of this method, the limitation lies in the small sample size. Furthermore, it is important to highlight that as the operational duration of Swift extends, the quality of its data gradually declines. When applying \cite{2012ApJ...749...68S}'s criteria to Swift GRBs up until December 2023, the redshift completeness reduces to 68\%. The second approach involves employing an incomplete sample while meticulously accounting for the impact of various selection effects.

In this paper, we utilize Swift GRB data up until December 2023 to isolate a purely Type II GRB sample, reconstruct its LF, and investigate the evolution of GRBs with redshift to distinguish between different models. In Chapter 2, we describe the sample and selection criteria. Chapter 3 details our methods for constructing the LF and the fitting process. Chapter 4 presents the results, followed by a brief discussion in Chapter 5. Throughout this paper, a flat $\Lambda$CDM cosmological model with $H_0 = 70 \, \text{km} \, \text{s}^{-1} \, \text{Mpc}^{-1}$, $\Omega_m = 0.3$, and $\Omega_\Lambda = 0.7$ is adopted.

\section{Samples} \label{Samples}

As of December 2023, Swift BAT has detected a total of 1612 GRBs\footnote{\url{https://swift.gsfc.nasa.gov/archive/grb_table/}}, among which 514 have associated redshift measurements\footnote{ See \url{https://swift.gsfc.nasa.gov/results/batgrbcat/summary_cflux/summary_general_info/GRBlist_redshift_BAT.txt} and the references listed in the table note for Table \ref{tab:gamma_bursts}.}
. The conventional classification based on $T_{90}=2s$  to distinguish between long and short GRBs tends to incorporate some Type I GRBs (or compact star GRBs) into the sample of LGRBs. Our objective is to identify a pure sample of Type II GRBs. \cite{2009ApJ...703.1696Z} proposed a stringent and well-defined criterion for differentiating between Type I and Type II GRBs. However, many of the criteria—such as the  supernova (SN) signature, specific star formation rate (SSFR) information, or the nature of the circumstellar medium—are often lacking observational data for many GRBs. For instance, with regard to SN signatures, only a small fraction of GRBs exhibit a clear SN signature, and these are predominantly observed at low redshifts \citep{2022ApJ...938...41D}. Consequently, this criterion is insufficient for constructing a complete sample or for determining the type  of every GRB. To more effectively identify a Type II sample, we employed the following three criteria:

\pagebreak 
\textbf{(1) Peak flux } $P \geq 1 \, \text{ph} \, \text{cm}^{-2} \, \text{s}^{-1}$

\begin{figure}
	\includegraphics[width=\columnwidth]{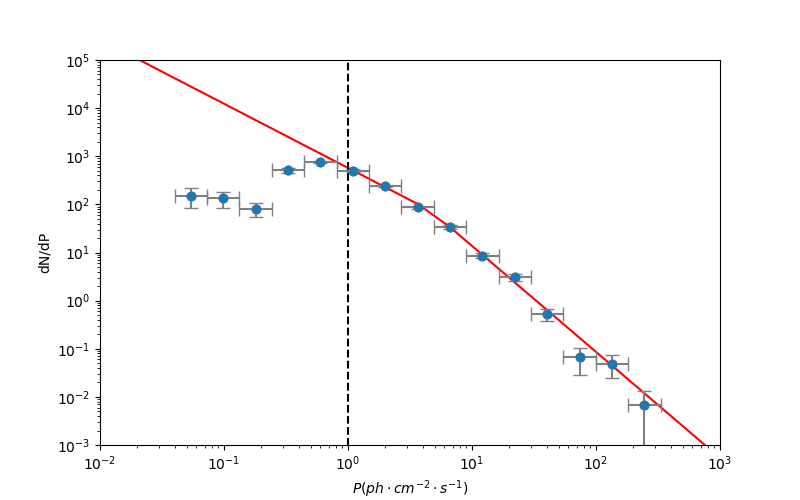}
    \caption{The peak-flux distribution for the 1612 GRBs recorded by Swift BAT. The red solid line represents the best-fit result to the observed peak-flux distribution for bursts with \(P \geq 1 \, \text{ph} \, \text{cm}^{-2} \, \text{s}^{-1}\), modeled using a BPL function. The dashed line indicates the flux threshold of \(1 \, \text{ph} \, \text{cm}^{-2} \, \text{s}^{-1}\), above which the effect of instrumental selection biases on the detection of fainter bursts is minimized.
 }
    \label{fig:1}
\end{figure}

One of the challenges in studying the LF of GRBs is accounting for selection effects. The Swift/BAT trigger system is complex, and its sensitivity to GRBs cannot be easily parameterized \citep{2006ApJ...644..378B}. Consequently, not all low-power GRBs with peak fluxes just above the instrument's threshold are successfully triggered. A practical approach to mitigating these selection effects is to increase the peak-flux threshold when selecting GRBs. Assume that the intrinsic peak flux distribution of GRBs follows a BPL. As shown in  figure \ref{fig:1}, for GRBs with peak fluxes $P \geq 1 \, \text{ph} \, \text{cm}^{-2} \, \text{s}^{-1}$, the observed peak flux distribution agrees well with the BPL model. However, for GRBs with $P < 1 \, \text{ph} \, \text{cm}^{-2} \, \text{s}^{-1}$, the  distribution deviates significantly from the BPL model. This deviation is likely caused by selection effects.  To mitigate the impact of selection effects on our results, we limit our sample to GRBs with $P \geq 1 \, \text{ph} \, \text{cm}^{-2} \, \text{s}^{-1}$ 

\textbf{(2) $T_{90,i} \geq 2 s$ }

$T_{90,i}$ represents the intrinsic duration of a GRB, defined as $T_{90}/(1 + z)$.  The bimodal distribution of GRBs, characterized by their $T_{90}$ duration into long and short bursts, was first observed by Konus-Wind \citep{1981Ap&SS..80....3M} and later corroborated by data from BATSE \citep{1993ApJ...413L.101K} and Swift \citep{2013ApJ...764..179B}.
Using $ T_{90} = 2 s$ to classify GRBs into two types is a qualitative rather than a quantitative method. This threshold is based on the minimum value at the trough between the two peaks of the bimodal $T_{90}$ distribution. For a more quantitative analysis of the intrinsic duration of GRBs, the GRB duration for an accretion-powered engine can be defined as \citep{2012ApJ...749..110B,2025JHEAp..45..325Z}
\begin{equation}
\begin{split}
T_{\text{GRB}} \simeq \max(t_{\text{ff}}, t_{\text{acc}}) - t_{\text{bo}}.
\end{split}
\end{equation}
Here, $t_{\text{ff}}$ represents the free-fall timescale of the progenitor star, which indicates the time required for the available material to fall onto the disk, such as at its outer boundary. $t_{\text{acc}}$ denotes the typical timescale for a mass element to migrate from the disk's outer boundary to the central object. Lastly, $t_{\text{bo}}$ corresponds to the time needed for the jet to penetrate and escape the envelope of the progenitor system.

For Type I GRBs, $t_{\text{ff}} \ll t_{\text{acc}} $, so $T_{\text{GRB}} \simeq t_{\text{acc}} - t_{\text{bo}}$. 
 With typical parameter values, one can estimate $T_{\text{GRB}} < 2\, \text{s}$ . In contrast, for Type II GRBs, $t_{\text{ff}} \gg t_{\text{acc}}$, so $T_{\text{GRB}} \simeq t_{\text{ff}} - t_{\text{bo}} $. 
 Since $t_{\text{ff}} \gg t_{\text{bo}}$ in most scenarios, the intrinsic duration of the GRB is primarily determined by $t_{\text{ff}}$.
 Consequently, the use of $T_{90,i}$ as a classification criterion to distinguish between Type I and Type II GRBs is not only reasonable and consistent with their respective physical timescales but also aligns with common practices in previous studies \citep{2009ApJ...703.1696Z,2010ApJ...725.1965L,2020MNRAS.492.1919M}.

On the other hand, many factors can affect the observed duration of GRBs, including background noise, viewing angle, and signal-to-noise ratio \citep{2013ApJ...765..116K,2022ApJ...927..157M}. These effects are particularly significant for faint GRBs. From Fig. 8 of \cite{2013ApJ...765..116K}, it can be seen that for low-luminosity, low-SNR GRBs, the observed duration can be as short as one-tenth of the intrinsic duration. However, for bright GRBs with a peak flux  $P \geq 1 \, \text{ph} \, \text{cm}^{-2} \, \text{s}^{-1}$, the observed duration, while still slightly shorter than the intrinsic duration, maintains a stable proportional relationship with the intrinsic duration. 
 Since our sample consists of GRBs with  Peak Flux  $P \geq 1 \, \text{ph} \, \text{cm}^{-2} \, \text{s}^{-1}$, $T_{90,i} $ serves as a reasonable proxy for the intrinsic duration of GRBs in this study.



\textbf{(3) Conforms to the Amati relation for Type II  GRBs}

\begin{figure}
	\includegraphics[width=\columnwidth]{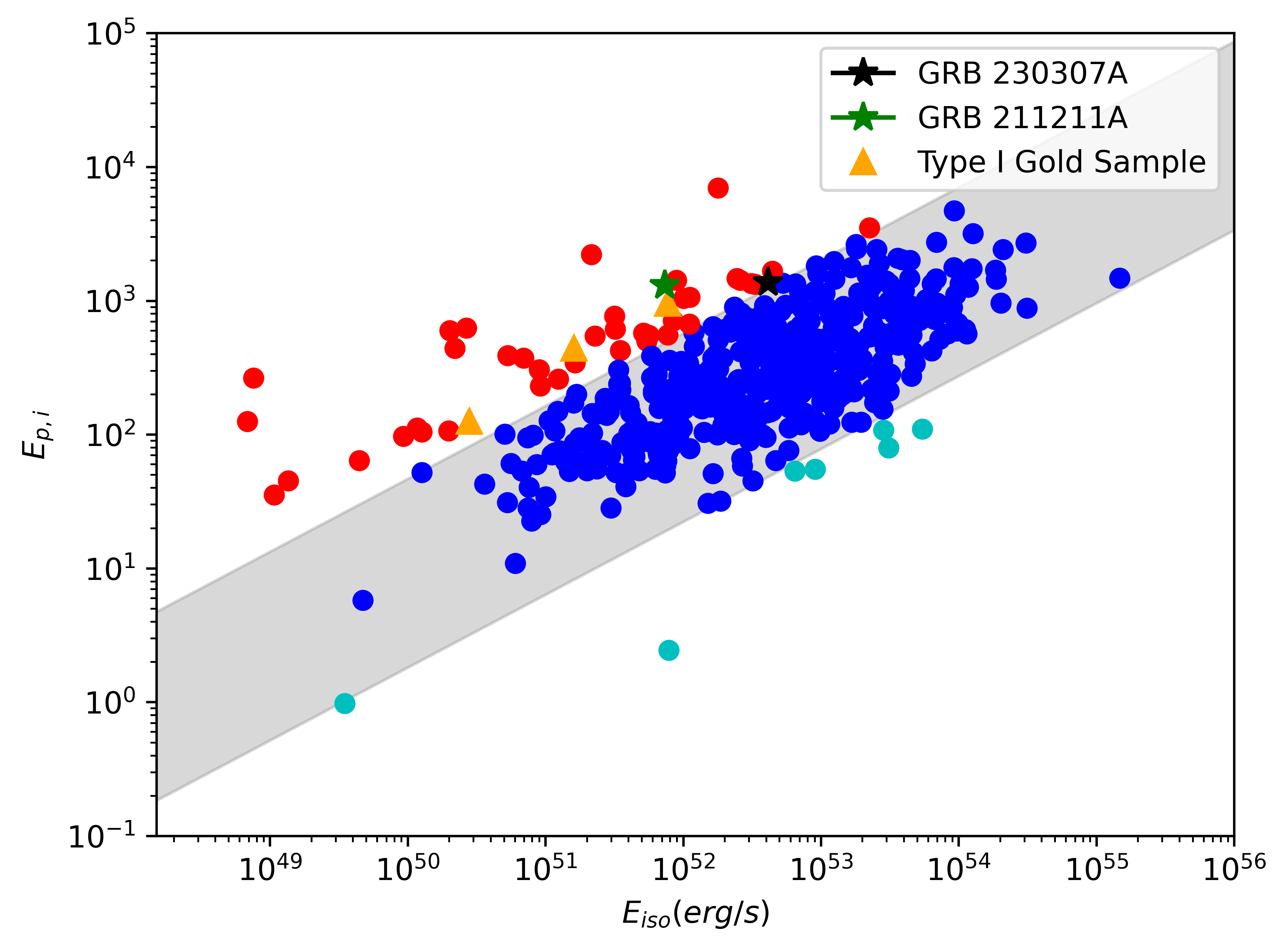}
    \caption{The relationship between $E_{iso}$ and $E_{p,i}$ for all Swift GRBs with $T_{90} \geq 2s$.  We applied the classic Amati relation to fit the $E_{iso}-E_{p,i}$ curve.  The shaded area represents the 1-sigma confidence interval. Sources within the shaded region are classified as type II GRBs. }
    \label{fig:2}
\end{figure}

The Amati relation \citep{2002A&A...390...81A,2006MNRAS.372..233A,2008MNRAS.391..577A} is one of the well-known empirical relationships in GRB studies, and serves as an important criterion for distinguishing between Type I and Type II GRBs \citep{2009ApJ...703.1696Z}. In this study, we used the Amati relation to identify Type II GRB samples. The \( \text{E}_{\text{pi}} \) and \( \text{E}_{\text{iso}} \) values of all LGRBs with known redshifts are shown in Figure \ref{fig:2}, where \( \text{E}_{\text{pi}} = \text{E}_{\text{p}} \times (1 + z) \) and \( \text{E}_{\text{iso}} \) can be obtained as follows:
\begin{equation}
\begin{split}
E_{iso} &= 4\pi D_{L}^{2}S_{bolo}(1+z)^{-1} \\
D_{L} &= \frac{(1+z)c}{H_{0}}\int_{0}^{z}\frac{dz^{\prime}}{\sqrt{\Omega_{M}(1+z^{\prime})^{3}+\Omega_{\Lambda}}} \\
S_{bolo} &= S_{\gamma}\frac{\int_{1/(1+z)}^{10000/(1+z)}E\phi(E)dE}{\int_{E_{min}}^{E_{max}}E\phi(E)dE}
\end{split}
\end{equation}

The distribution presented in Figure \ref{fig:2} exhibits somewhat broader dispersion compared to previous Type II Amati relations (e.g., \cite{2022MNRAS.516.2575J}). This broader dispersion is primarily attributed to the narrow energy band (15-150 keV) of Swift/BAT data, which restricts its ability to accurately constrain the high-energy spectral components of GRBs \citep{2005SSRv..120..143B,2011ApJS..195....2S}. Consequently, only approximately 200 GRBs have spectra characterized by Band function fitting \citep{1993ApJ...413..281B}. For GRBs where Band spectral fitting was unavailable, $\text{E}_{\text{p}}$  values were adopted from CPL model fits reported in previous studies. While simpler and easier to apply, the CPL model often results in less precise constraints on  $\text{E}{\text{p}} $, leading to a more dispersed  $E_{p} - E_{iso}$  relationship. For our luminosity function analysis, it was necessary to use a sample as large as possible. As a result, our sample includes both GRBs with Band spectral fits and those with CPL fits. While this combination ensures a sufficiently large dataset, the inclusion of CPL-fitted GRBs introduces broader scatter into the Type II Amati relations, as shown in Fig. \ref{fig:2}.




The shaded region in Figure \ref{fig:2} represents the 1-sigma dispersion  range of the best-fit model, where $\sigma$ refers to the standard deviation of the residuals ($\sigma_\text{residuals}$) between the observed data points and the best-fit line.
Notably, our sample contains  three GRBs (GRB 050724, GRB 060614, GRB 061006) from the Type I Gold sample identified by \cite{2009ApJ...703.1696Z}, all of which are robustly excluded by our selection criteria. These GRBs are  marked with triangular symbols in Figure \ref{fig:2}. Additionally, two LGRBs with substantial evidence of being associated with compact binary mergers (GRB 211211A and GRB 230307A) are also indicated in Figure \ref{fig:2}. GRB 211211A (\( T_{90} > 30 \text{s}, z = 0.076 \)) and GRB 230307A (\( z = 0.0646, T_{90} \sim 35 \text{s} \)) were not observed by Swift, and thus are not included in our sample; nevertheless, they are effectively excluded by our selection criteria.

Notably, A total of seven GRBs situated below the shaded region in Figure \ref{fig:2} include two GRBs (211024B, 171205A) where the \( \text{E}_{\text{p}} \) values were inferred using the empirical Ep–L correlation, owing to the absence of direct spectral measurements. Moreover, two additional GRBs (140301A, 141026A) were fitted with Band spectra but lacked the \( \alpha \) parameter, necessitating the use of the default value of \( -1 \) \citep{2021MNRAS.508...52L}. The remaining three GRBs were modeled using CPL spectra instead of Band spectra. The observed deviations of these GRBs below the shaded region in Figure \ref{fig:2} can likely be attributed to the limitations in the quality of the available data for these events.

The final sample comprises 307 Swift GRBs that meet the above selection criteria. The names of these GRBs along with the  references are listed in Table \ref{tab:gamma_bursts}.

\section{ANALYSIS METHOD} \label{sec:METHOD}

The Maximum Likelihood method serves as a powerful tool for determining the optimal parameters in models with multiple variables \citep{1983ApJ26935M,2006ApJ...643...81N,2009ApJ...699..603A,2012ApJ...751..108A,2010ApJ...720..435A,2014MNRAS.441.1760Z,2016MNRAS.462.3094Z,2019MNRAS.490..758Q,2019MNRAS.488.4607L,2021MNRAS.508...52L,2022ApJ...938..129L}. For our purposes, the likelihood function can be written as:

\begin{equation}
    \mathcal{L} = \exp(-N_{\text{exp}}) \prod_{i=1}^{N_{\text{obs}}} \Phi(L_i, z_i, t_i)
	\label{eq:L}
\end{equation}
where \(N_{\text{exp}}\) represents the expected number of GRBs, \(N_{\text{obs}}\) denotes the observed sample count, and \(\Phi(L, z, t)\) is the observed rate of bursts per unit time at redshift \(z \sim z + dz\) with luminosity \(L \sim L + dL\), given by:

\begin{equation}
\begin{split}
    \Phi(L, z, t) &= \frac{d^{3} N}{dtdzdL} \\
    &= \frac{d^{3}N}{dtdVdL} \times \frac{dV}{dz} \\
    &= \frac{\Delta \Omega}{4\pi} \theta(P) \frac{\psi(z)}{(1+z)} \phi(L,z) \times \frac{dV}{dz}
\end{split}
\label{eq:Phi}
\end{equation}
Here, \(\Delta \Omega = 1.4 \, \text{sr}\) (half-coded) \citep{2005SSRv..120..143B} represents the field of view of Swift/BAT. \(\theta(P)\equiv\theta_{\gamma}(P)\theta_{z}(P)\) is the detection efficiency, which accounts for the likelihood of triggering and measuring the redshift for a burst with peak flux \(P\). Our sample is defined by a selection criterion of $P \geq 1 \, \text{ph} \, \text{cm}^{-2} \, \text{s}^{-1}$, under which we adopt the assumption $\theta_{\gamma}(P) = 1$. For $\theta_{z}(P)$, we apply the model presented by \cite{2021MNRAS.508...52L}, where $\theta_{z}(P)$ is expressed as $\theta_{z}(P) = \frac{1}{1 + (2.09 \pm 0.26) \times (0.96 \pm 0.01)^P}$.

The function \(\psi(z)\) describes the comoving event rate of GRBs (in units of \( \text{Mpc}^{-3} \, \text{yr}^{-1} \)) as a function of redshift \(z\), with the factor \((1 + z)^{-1}\) accounting for cosmological time dilation. The term \(\phi(L, z)\) is the normalized GRB LF, which may evolve with redshift depending on the model. The comoving volume element in a flat \(\Lambda\)CDM model is given by \( \frac{dV(z)}{dz} = \frac{4\pi c D_{L}^{2}(z)}{H_{0}(1 + z)^2 \sqrt{\Omega_{m}(1 + z)^3 + \Omega_{\Lambda}}} \), where \(D_L(z)\) is the luminosity distance at redshift \(z\).

Within the collapsar model framework, each GRB signals the death of a massive star, suggesting that the GRB formation rate \(\psi(z)\) is inherently tied to the SFR, \(\psi_{\star}(z)\), where \(\psi(z) = \eta \psi_{\star}(z)\). Here, \(\eta\) represents the GRB formation efficiency. The SFR \(\psi_{\star}(z)\), in unit of \(M_{\odot} \, \text{yr}^{-1} \, \text{Mpc}^{-3}\), can be approximated as \citep{2006ApJ...651..142H,2008MNRAS.388.1487L}:

\begin{equation}
    \psi_{\star}(z) = \frac{0.0157 + 0.118z}{1 + (z/3.23)^{4.66}}
\end{equation}

For the GRB LF \(\phi(L, z)\), we use a BPL:

\begin{equation}
\phi(L,z)=\frac{A}{\ln(10)L}\left\{
\begin{array}{ll}
(\frac{L}{L_{c}(z)})^{a}; & L\leq L_{c1}(z) \\
   \\
(\frac{L}{L_{c}(z)})^{b}; & L>L_{c}(z)
\end{array}
\right.
\end{equation}
where \(A\) is a normalization constant. \(a\) and \(b\) are the power-law indices before and after the break luminosity \(L_{c}\). If this BPL model does not fit the data well, a TPL form is considered:

\begin{equation}
\phi(L,z)=\frac{A}{\ln(10)L}\left\{
\begin{array}{ll}
(\frac{L}{L_{c1}(z)})^{a}; & L\leq L_{c1}(z) \\
                                              \\
(\frac{L}{L_{c1}(z)})^{b}; & L_{c1}(z) < L\leq L_{c2}(z) \\
                                                        \\
(\frac{L_{c2}(z)}{L_{c1}(z)})^{b}(\frac{L}{L_{c2}(z)})^{c}; &   L>L_{c2}(z)
\end{array}
\right.
\end{equation}
where \(a\), \(b\), and \(c\) are the power-law indices for the three segments, with \(L_{c1}\) and \(L_{c2}\) as the break luminosities.

Given the flux threshold for \(100\%\) trigger efficiency (i.e., \(P_{\text{lim}} = 1 \, \text{ph} \, \text{cm}^{-2} \, \text{s}^{-1}\) in the \(15-150 \, \text{keV}\) energy band), the expected number of GRBs can be expressed as:

\begin{equation}
\begin{split}
    N_{\text{exp}} = \frac{\Delta \Omega T}{4 \pi}\int_{0}^{z_{\text{max}}}\int_{\max[L_{\text{min}},L_{\text{lim}}(z)]}^{L_{\text{max}}} \theta(P(L,z)) \frac{\psi(z)}{1+z} \times \phi(L,z) \, dL \, dV(z)
\end{split}
\label{eq:Nexp}
\end{equation}

The duration of Swift's mission, encompassing the time span used in our analysis, is approximately 19 years, thus $T \approx 19 yr.$  Considering the current BAT sample, which includes data up to a redshift of \( z < 10 \), we set the maximum redshift for our analysis as \( z_{\text{max}} = 10 \). The LF is assumed to extend from \( L_{\text{min}} = 10^{49} \, \text{erg} \cdot \text{s}^{-1} \) to \( L_{\text{max}} = 10^{55} \, \text{erg} \cdot \text{s}^{-1} \) \citep{2015MNRAS.447.1911P}. The luminosity threshold defined in Equation \ref{eq:Nexp} is written as follows:

\begin{align}
   L_{\text{lim}}(z) = 4\pi D_{L}^{2}(z)P_{\text{lim}} \frac{\int_{1/(1+z) \, \text{keV}}^{10^{4}/(1+z) \, \text{keV}} EN(E) dE}{\int_{15 \, \text{keV}}^{150 \, \text{keV}} N(E) dE}
\end{align}

In this context, \( N(E) \) represents the photon spectrum of GRBs. We adopt a typical Band function spectrum \citep{1993ApJ...413..281B,2006ApJS..166..298K}, characterized by low-energy and high-energy spectral indices of \( -1 \) and \( -2.3 \), respectively. To broadly estimate the spectral peak energy \( E_{p} \) for a given luminosity \( L \), we adopt the empirical relation \( E_{p}-L \) \citep{2004ApJ...609..935Y,2012MNRAS.421.1256N}:

\[
\log [E_p(1 + z)] = -25.33 + 0.53 \log L
\]

To account for potential biases in the connection between the GRB formation rate and the SFR, as well as the possibility of evolution in the GRB LF with redshift, we introduce an additional evolution factor \( (1 + z)^\delta \), where \( \delta \) is a free parameter. We explore three distinct models:

1. The GRB formation rate strictly follows the SFR, \( \psi(z) = \eta \psi_{\star}(z) \), and the LF does not evolve with redshift \( L_{ci}(z) = L_{ci,0} = \text{constant} \).

2. The GRB formation rate scales with the SFR, but the break luminosity in the GRB LF increases with redshift \( L_{ci}(z) = L_{ci,0}(1 + z)^\delta \).

3. The formation rate follows the SFR and includes an additional evolution factor \( \psi(z) = \eta \psi_{\star}(z)(1 + z)^\delta \), while the LF remains unevolved.

For each model, we optimize the free parameters by maximizing the likelihood function (Equation \ref{eq:L}). Given the complexity of our models, we employ Markov Chain Monte Carlo (MCMC) sampling technique to determine the best-fit values of model parameters and associated \( 1 \sigma \) uncertainty. The MCMC code EMCEE is used in this analysis \citep{2013PASP..125..306F}.

\begin{figure}[ht]
    \centering
    \begin{minipage}[b]{0.45\textwidth}
        \centering
        \includegraphics[width=\textwidth]{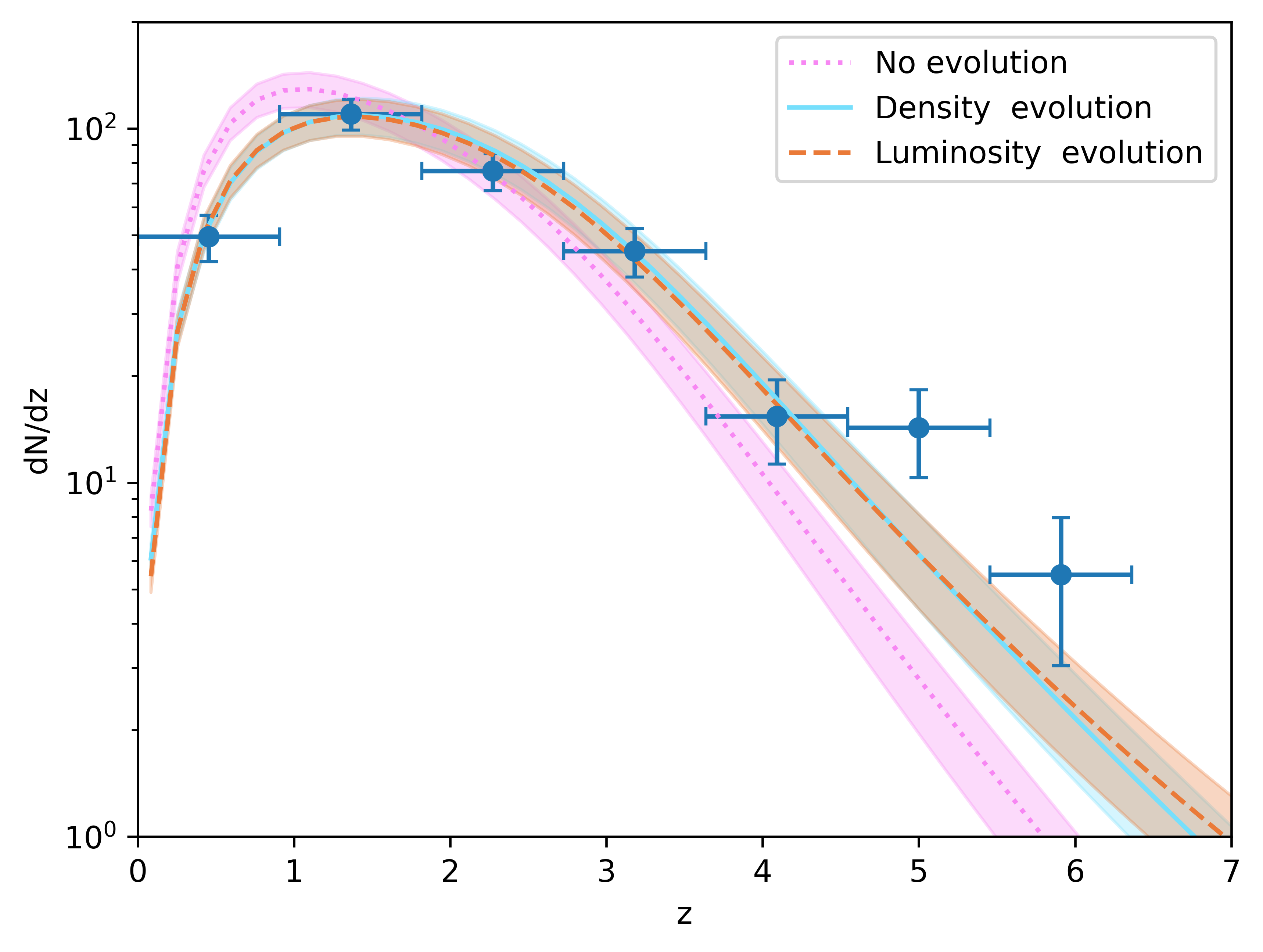}
    \end{minipage}
    \hspace{0.05\textwidth}
    \begin{minipage}[b]{0.45\textwidth}
        \centering
        \includegraphics[width=\textwidth]{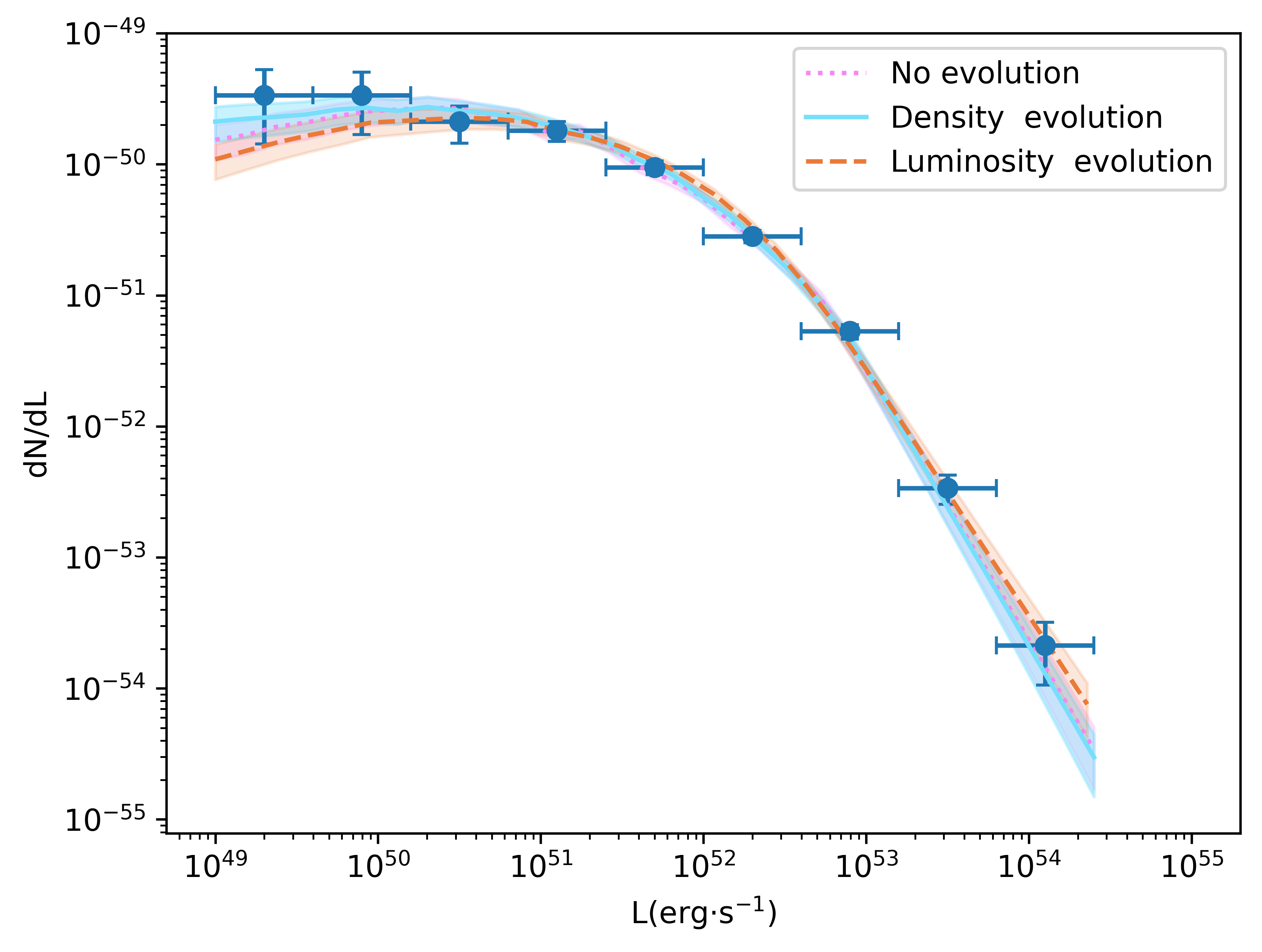}
    \end{minipage}
    \caption{Redshift and luminosity distributions of our sample (steel blue solid points) accompanied by Poisson error bars. The different curves correspond to the predicted distributions from various best-fit models: the no evolution model (pink dotted lines), the density evolution model (pastel blue solid lines), and the luminosity evolution model (orange dashed lines). The shaded regions represent the 1$\sigma$ confidence intervals for each respective model. A BPL LF  is assumed across all models.}

    \label{fig:PL}
\end{figure}

\begin{figure}[ht]
    \centering
    \begin{minipage}[b]{0.45\textwidth}
        \centering
        \includegraphics[width=\textwidth]{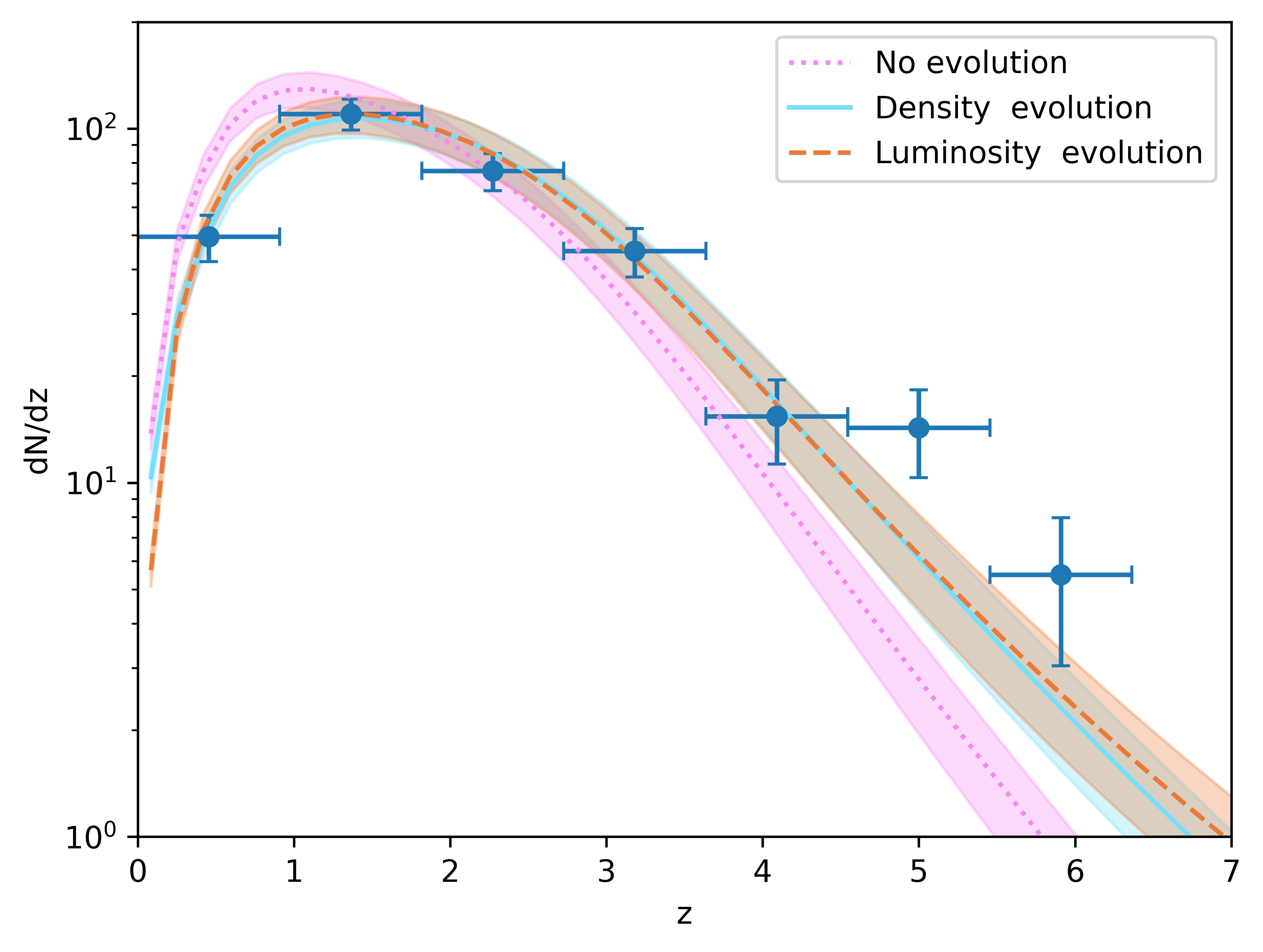}
    \end{minipage}
    \hspace{0.05\textwidth}
    \begin{minipage}[b]{0.45\textwidth}
        \centering
        \includegraphics[width=\textwidth]{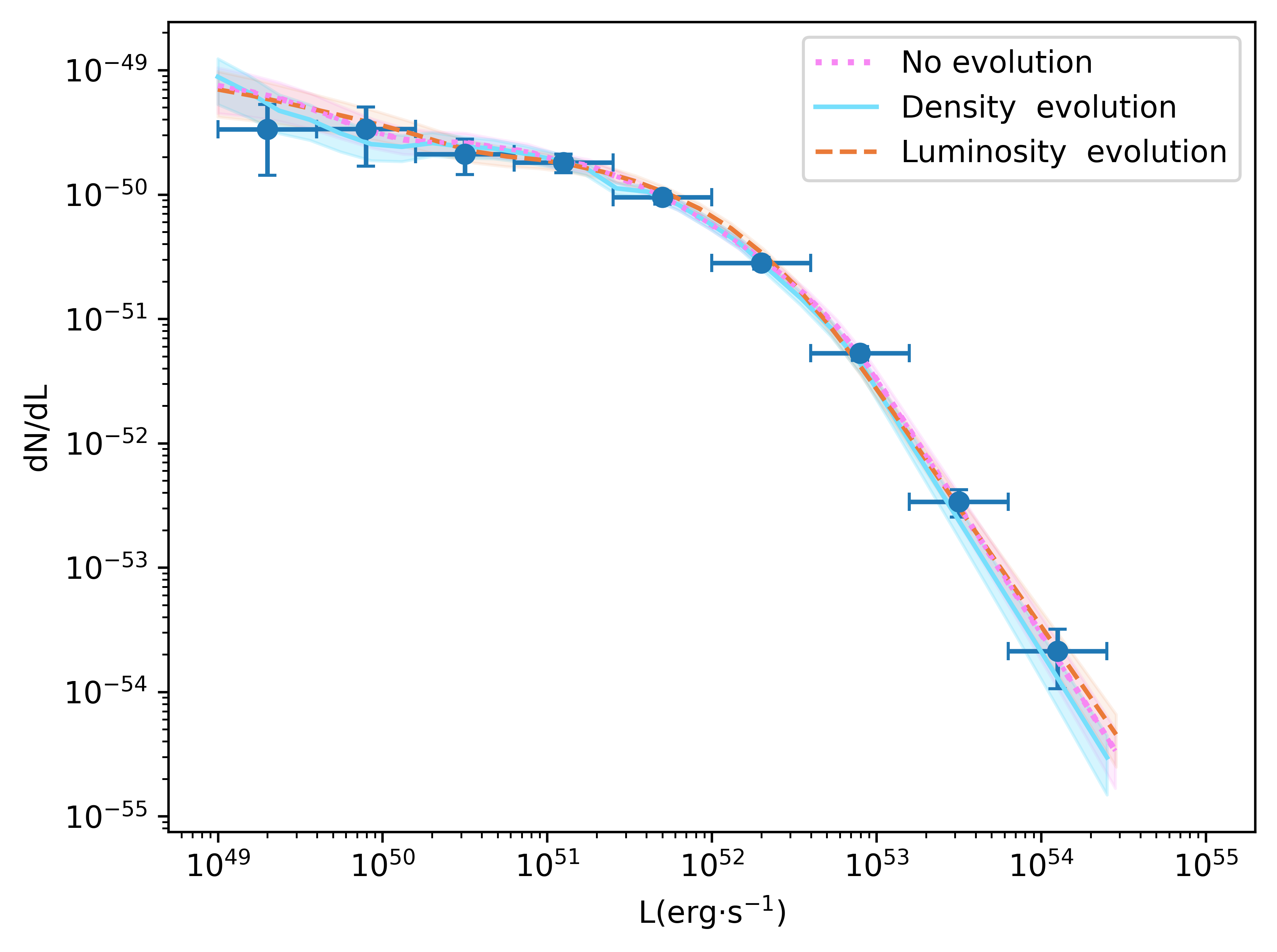}
    \end{minipage}
    \caption{Same as Figure \ref{fig:PL}, except now for the scenario with the assumed TPL LF.}
    \label{fig:TPL}
\end{figure}

\begin{table}
\centering
\caption{}\label{table1}
\begin{tabular}{@{}cccccccc@{}}
\toprule
Model & Evolution parameter  & $a$ & $b$ & $c$ & log $L_c$  & ln $L$ & AIC \\ 
 &  &   & & &  (erg s$^{-1}$) & & \\ \midrule
\multicolumn{8}{c}{Broken power-law luminosity function} \\\midrule
No evolution &… & $-0.34^{+0.03}_{-0.03}$ & $-1.24^{+0.27}_{-0.27}$ &… & $52.78^{+0.11}_{-0.12}$ & -150.46 & 308.92 \\

Luminosity evolution & $\delta = 1.74^{+0.24}_{-0.22}$  & $-0.36^{+0.07}_{-0.07}$ & $-1.08^{+0.19}_{-0.16}$ & …& $51.55^{+0.18}_{-0.19}$ & -127.52 & 265.04 \\

Density evolution & $\delta = 1.36^{+0.21}_{-0.20}$  & $-0.56^{+0.06}_{-0.06}$ & $-1.31^{+0.26}_{-0.27}$ &… & $52.91^{+0.12}_{-0.13}$ & -128.11 & 266.22 \\ \midrule
\multicolumn{8}{c}{Triple power-law luminosity function} \\ \midrule

No evolution &…  & $-0.99^{+0.75}_{-0.70}$ & $-0.33^{+0.05}_{-0.05}$ & $-1.24^{+0.23}_{-0.25}$ & $50.16^{+0.36}_{-0.41}, 52.76^{+0.08}_{-0.09}$ & -149.67 & 311.34 \\
Luminosity evolution & $\delta = 1.79^{+0.24}_{-0.36}$  & $-1.01^{+0.54}_{-0.61}$ & $-0.32^{+0.11}_{-0.12}$ & $-1.11^{+0.13}_{-0.19}$ & $50.24^{+0.35}_{-0.44}, 51.53^{+0.17}_{-0.28}$ & -126.32 & 266.64 \\

Density evolution & $\delta = 1.34^{+0.31}_{-0.32}$  & $-1.29^{+0.18}_{-0.23}$ & $-0.55^{+0.08}_{-0.09}$ & $-1.31^{+0.22}_{-0.21}$ & $49.94^{+0.19}_{-0.22}, 52.91^{+0.10}_{-0.11}$ & -127.65 & 269.30 \\

\bottomrule
\end{tabular}
\begin{minipage}{\textwidth}
\footnotesize
\textbf{Note.} The parameter values were obtained as the medians of the most optimal parameters from the Monte Carlo sample, with uncertainties denoted by the 68\% confidence intervals surrounding these median values.

\end{minipage}
\end{table}

\section{RESULTS} \label{sec:RaC}

Utilizing the aforementioned analytical approach, we fine-tuned the model's free parameters, including the GRB LF, the evolution parameter, and the GRB formation efficiency ($\eta$). The optimal parameters, along with their corresponding $1\sigma$ confidence intervals for various models, are summarized in Table \ref{table1}. To assess which model is statistically favored by the observational data, we provide the log-likelihood value $\ln L$ and the Akaike Information Criterion (AIC) score in the last two columns of Table \ref{table1}. The AIC score for each model fit is computed as $\text{AIC} = -2 \ln L + 2n$, where $n$ denotes the number of free parameters \citep{1974ITAC...19..716A, 2007MNRAS.377L..74L}. With $AIC_i$ representing model $M_i$, the unnormalized confidence that this model is accurate can be expressed as the Akaike weight $\exp(-AIC_i/2)$. In a comparative analysis, model \(M_i\) demonstrates a relative probability of being the accurate model:

\begin{align}
P(\mathcal{M} _{i})=\frac{\exp(-AIC_{i}/2)}{\exp(-AIC_{1}/2)+\exp(-AIC_{2}/2)}
\end{align}

 Accordingly, the difference \( \text{AIC}_2 - \text{AIC}_1 \) quantifies the preference for model \( M_1 \) over model \( M_2 \). Figures \ref{fig:PL} and \ref{fig:TPL} depict the redshift and luminosity distributions for GRBs with photon fluxes \( P \geq 1 \) ph cm\(^{-2}\) s\(^{-1}\). For each model, we evaluated two distinct GRB LF formulations as outlined in Section \ref{sec:METHOD}. Figure \ref{fig:PL} showcases the best-fitting results for different models utilizing the BPL LF, whereas Figure \ref{fig:TPL} presents the outcomes using the TPL LF.

\subsection{Exclusion of the No-Evolution Model}

In the no-evolution scenario, Type II GRBs are assumed to trace the SFR without any redshift evolution in their LF, i.e., \( L_c(z) = L_c,0 \) remains constant. Figure \ref{fig:PL} depicts the predicted redshift and luminosity distributions across various models, based on an assumed BPL LF. The results indicate that the no-evolution model (represented by the pink dotted lines) fails to adequately reproduce the observed distributions within our dataset. Specifically, the peak of the predicted redshift distribution occurs at a lower redshift than observed, and the model significantly underestimates both the GRB rate at high redshifts and the low-luminosity end of the distribution. These findings are consistent with those of \cite{2021MNRAS.508...52L}, derived from LGRBs. 
Applying the Akaike Information Criterion (AIC), this model can be decisively ruled out, with a likelihood of approximately \( 3 \times 10^{-10} \) compared to the luminosity evolution model using the BPL LF.

Figure \ref{fig:TPL}, which presents analogous data but employs a TPL LF, shows that while the fit to the luminosity distribution improves, the predicted redshift distribution from the no-evolution model remains inconsistent with the observational data. Using the AIC criterion, the no-evolution model is further excluded, with an estimated probability of only \( 2 \times 10^{-10} \) of being the correct model in comparison to the luminosity evolution model incorporating the TPL LF.
\subsection{Indistinguishability of the Luminosity and Density Evolution Models}
We explore the implications of a redshift-dependent GRB LF on the detection rates of GRBs at high redshifts. Assuming that the GRB formation rate closely follows the cosmic SFR, we propose that the characteristic luminosity, $L_c(z)$, evolves with redshift according to $L_c(z) = L_{c,0}(1 + z)^\delta$. This model suggests that GRBs observed at higher redshifts are intrinsically more luminous compared to those at lower redshifts. By employing a BPL LF, our analysis reveals that a significant luminosity evolution with $\delta = 1.74^{+0.24}_{-0.22}$ provides an excellent fit to the observed redshift and luminosity distributions, as demonstrated by the orange dashed lines in Figure \ref{fig:PL}.

A significant rise in the formation rate of GRBs as a function of redshift can lead to an increased detection of high-redshift GRBs. Within this context, while the break luminosity of the GRB LF remains constant, the GRB occurrence rate is intrinsically tied to the cosmic SFR, further modulated by an evolutionary factor denoted as $(1 + z)^\delta$. This relationship is mathematically expressed as $\psi(z) = \eta\psi_\ast(z)(1 + z)^\delta$. Employing a BPL LF, our analysis indicates that a significant density evolution, characterized by $\delta = 1.36^{+0.21}_{-0.20}$, provides the best agreement with the observed data, as depicted by the pastel blue solid lines in Figure \ref{fig:PL}.

The analysis presented in Table \ref{table1} demonstrates that both the Luminosity Evolution and Density Evolution models offer a strong fit to the data, with their respective AIC values being nearly indistinguishable. This close alignment underscores the challenge in directly differentiating between the two models. Such difficulties in model differentiation have been noted in previous studies, such as those by \cite{2012ApJ...749...68S} and \cite{2021MNRAS.508...52L}. \cite{2012ApJ...749...68S} examined a dataset of 58 LGRBs with high redshift completeness and  peak flux $P \geq 2.6 \ \text{ph cm}^{-2} \ \text{s}^{-1}$, while \cite{2021MNRAS.508...52L} analyzed a larger sample of 302 LGRBs with peak flux $P \geq 1 \ \text{ph cm}^{-2} \ \text{s}^{-1}$. However, with the availability of higher-quality observational data from upcoming missions like SVOM \citep{2016arXiv161006892W} and the use of more advanced analytical techniques, future research holds the potential to clarify the distinctions between these models.

\subsection{Triple Power Law vs Broken Power Law}

Previous studies, including those by \cite{2015ApJ...812...33S} and \cite{2021MNRAS.508...52L}, have suggested that the TPL model provides a superior fit to the LF of LGRBs compared to the BPL model. However, our focused analysis on Type II GRBs indicates that the maximum likelihood estimates of the TPL and BPL models are nearly indistinguishable, showing only a slight preference for the TPL model. Importantly, when model complexity is taken into account using the Akaike Information Criterion (AIC), the TPL model yields a higher AIC value than the BPL, suggesting a less favorable balance between model fit and complexity. These results imply that the previously reported superiority of the TPL model may have been influenced by the inclusion of non-Type II GRB samples. A thorough discussion  will be presented in the subsequent section.

\section{CONCLUSIONS and Discussion} \label{sec:Discussion}
We analyzed a dataset of Swift GRBs observed from 2004 to 2023, focusing on Type II GRBs. In their seminal 2009 study, \cite{2009ApJ...703.1696Z} developed a robust and precise framework for differentiating between Type I and Type II GRBs. However, the implementation of these classification criteria—such as the detection of a supernova (SN) signature, the SSFR, or the properties of the circumstellar medium—remains challenging due to the limited observational data available for many GRBs.
Given that a larger sample is conducive to constructing the LF, we employed two selection criteria commonly used in GRB samples: an intrinsic duration of \( T_{90,i} \geq 2 \, \text{s} \) and consistency with the Amati relation characteristic of Type II GRBs, which is also one of the critical selection criteria mentioned by \cite{2009ApJ...703.1696Z}. To mitigate the influence of selection effects on the results, we introduced a third selection criterion: a peak flux threshold of \( P \geq 1 \, \text{ph} \, \text{cm}^{-2} \, \text{s}^{-1} \). GRBs with a peak flux below this threshold exhibit a distribution that significantly deviates from the BPL  distribution.

 This rigorous selection process yielded a sample of 307 GRBs. Notably, three Type I Gold Sample GRBs \citep{2009ApJ...703.1696Z} (GRB 050724, GRB 060614, and GRB 061006) and two GRBs with strong evidence suggesting origins from  compact star mergers (GRB 211211A and GRB 230307A) were excluded by our criteria.

We explored three scenarios: no evolution, luminosity evolution, and density evolution. We constructed LFs using both BPL and TPL models. Our main findings are as follows:

1) The no-evolution model is decisively ruled out. The AIC model selection criterion indicates that the no-evolution model can be rejected with a very low probability (approximately $\sim 10^{-10}$) compared to the luminosity evolution model assuming a BPL LF. Additionally, the no-evolution model fails to accurately represent the observed distributions in our sample, predicting a peak in the redshift distribution at a lower value than observed and underestimating the rate of GRBs at high redshifts.

2) Both luminosity evolution and density evolution models provided satisfactory fits to the data. By applying the BPL LF, we ascertain that a pronounced luminosity evolution with $\delta = 1.74^{+0.24}_{-0.22}$ or a pronounced density evolution with $\delta = 1.36^{+0.21}_{-0.20}$ successfully matches the observed redshift and luminosity distributions. Both models yield comparable goodness-of-fit metrics, making it challenging to distinguish between the two. This outcome aligns with the findings of \cite{2012ApJ...749...68S}  and \cite{2021MNRAS.508...52L} for LGRBs.

3) According to the Akaike Information Criterion (AIC), the BPL model marginally outperforms the TPL model, contrasting with the conclusions of \cite{2015ApJ...812...33S} and \cite{2021MNRAS.508...52L}.

Interestingly, while the LGRBs tends to favor the TPL LF, our analysis suggests a preference for the BPL LF among Type II GRBs. This discrepancy might be explained by the existence of two distinct populations within the LGRB category: one corresponding to Type II GRBs, characterized by a BPL LF, and another consisting of non-Type II GRBs, also described by a BPL LF but with generally lower luminosity. The aggregation of these two BPL distributions could manifest as a TPL  LF when considering the entire LGRB population. Supporting evidence for this hypothesis includes:1) There are strong indications that GRB 211211A \citep{2022Natur.612..232Y} and GRB 230307A \citep{2024Natur.626..737L} likely originated from double compact star mergers, with GRB 170228A \citep{2024arXiv240702376W} as another potential candidate. 2) \cite{2023ApJ...958...37D} classified LGRBs into high-luminosity and low-luminosity groups, showing that the event rate of high-luminosity GRBs closely tracks the star formation rate, whereas the event rate of low-luminosity GRBs diverges significantly, suggesting different origins for these two classes. 3) \cite{2024ApJ...963L..12P} demonstrated that subtracting the SFR from the LGRB event rate results in a curve closely resembling the event rate of  compact star mergers, further implying multiple origins for LGRBs.

Clarifying the proportion of LGRBs that belong to the Type II category and identifying the origins of those that do not—whether due to  compact star mergers or other processes—is essential for understanding the relationship between GRBs and gravitational waves, the correlation between GRB event rates and the SFR, and the broader cosmological implications of GRBs. We anticipate that forthcoming observatories such as SVOM and Gcam will provide significant insights.

\begin{acknowledgments}
We sincerely thank Lan, GX for the valuable discussions and suggestions during the preparation of this manuscript. Additionally, we acknowledge the support from the Shandong Provincial Natural Science Foundation (Grant No. ZR2021MA021).

\end{acknowledgments}

%

\vspace{5mm}
\facilities{ Swift(XRT and BAT)}


\software{astropy\citep{2013A&A...558A..33A,2018AJ....156..123A,astropy:2022}, Emcee\citep{2013PASP..125..306F} }

\begin{longtable}{l c| l c |l c| l c |l c |l c}
\caption{} \label{tab:gamma_bursts} \\

\hline
GRB & Ref & GRB & Ref & GRB & Ref & GRB & Ref & GRB & Ref & GRB & Ref \\
\hline
\endfirsthead

\hline
GRB & Ref & GRB & Ref & GRB & Ref & GRB & Ref & GRB & Ref & GRB & Ref \\
\hline
\endhead

\endfoot

\hline
\endlastfoot
041228	&	3			&	061121	&	30			&	081008	&	31			&	110818A	&	1	,	3	&	140430A	&	1	,	4	&	180314A	&	2	,	16	\\
050219A	&	1	,	3	&	061126	&	30			&	081102	&	3			&	111008A	&	1	,	3	&	140506A	&	1	,	3	&	180325A	&	1	,	17	\\
050315	&	1	,	3	&	061202	&	1	,	4	&	081109A	&	1	,	3	&	111107A	&	1	,	3	&	140509A	&	1	,	4	&	180329B	&	1	,	18	\\
050318	&	30			&	061222A	&	2	,	3	&	081121	&	31			&	111228A	&	1	,	3	&	140512A	&	1	,	3	&	180404A	&	1			\\
050319	&	1	,	4	&	061222B	&	1	,	3	&	081203A	&	2	,	3	&	111229A	&	1	,	4	&	140518A	&	1	,	3	&	180510B	&	1			\\
050401	&	30			&	070103	&	1	,	4	&	081210	&	1	,	3	&	120118B	&	1	,	3	&	140629A	&	1	,	3	&	180620B	&	1	,	19	\\
050410	&	3			&	070306	&	1	,	4	&	081221	&	32			&	120119A	&	1	,	3	&	140703A	&	1	,	3	&	180624A	&	1	,	4	\\
050502B	&	1	,	3	&	070318	&	1	,	4	&	081222	&	31			&	120326A	&	1	,	3	&	140710A	&	1	,	3	&	180720B	&	1	,	2	\\
050505	&	1	,	4	&	070328	&	1	,	4	&	090102	&	31			&	120327A	&	1	,	3	&	140907A	&	1	,	5	&	180728A	&	1	,	20	\\
050525A	&	2	,	3	&	070419B	&	1	,	4	&	090113	&	1	,	4	&	120404A	&	1	,	4	&	141109A	&	1	,	4	&	181020A	&	1	,	21	\\
050603	&	30			&	070420	&	1	,	2	&	090201	&	1	,	3	&	120521C	&	1	,	4	&	141220A	&	1	,	3	&	181110A	&	1	,	22	\\
050716	&	3			&	070508	&	3			&	090401B	&	3			&	120624B	&	1	,	2	&	141221A	&	1	,	6	&	181213A	&	1	,	2	\\
050801	&	1	,	3	&	070521	&	32			&	090404	&	1	,	3	&	120712A	&	1	,	3	&	141225A	&	1	,	4	&	190106A	&	1	,	23	\\
050802	&	1	,	4	&	070529	&	1	,	3	&	090418A	&	2	,	3	&	120722A	&	1	,	4	&	150120B	&	1	,	7	&	190114C	&	1	,	24	\\
050820A	&	2	,	3	&	070612A	&	2	,	3	&	090424	&	31			&	120729A	&	1	,	3	&	150206A	&	1	,	4	&	190324A	&	2	,	25	\\
050822	&	1	,	4	&	070621	&	3			&	090516A	&	2	,	3	&	120802A	&	1	,	3	&	150301B	&	1	,	4	&	190613A	&	1	,	26	\\
050922B	&	1	,	3	&	070721B	&	1	,	3	&	090530	&	1	,	3	&	120811C	&	32			&	150314A	&	1	,	4	&	190719C	&	1	,	27	\\
051008	&	1	,	4	&	070808	&	3			&	090618	&	32			&	120815A	&	1	,	4	&	150323A	&	1	,	4	&	191004B	&	2	,	28	\\
051016B	&	1	,	3	&	070810A	&	2	,	3	&	090709A	&	3			&	120907A	&	1	,	3	&	150403A	&	1	,	4	&	191011A	&	2	,	29	\\
051109A	&	30			&	071003	&	31			&	090715B	&	33			&	120909A	&	2	,	3	&	150413A	&	1	,	4	&	191019A	&	2			\\
051111	&	1	,	3	&	071010B	&	1	,	3	&	090812	&	33			&	120922A	&	1	,	3	&	150424A	&	1	,	2	&	191221B	&	1	,	2	\\
060110	&	3			&	071025	&	1	,	3	&	090904B	&	3			&	121024A	&	1	,	4	&	150727A	&	1	,	4	&	200205B	&	1	,	2	\\
060111A	&	1	,	3	&	071028B	&	2	,	3	&	090926B	&	1	,	3	&	121027A	&	1	,	4	&	150818A	&	1	,	4	&	200528A	&	1	,	2	\\
060111B	&	3			&	071112C	&	1	,	3	&	091018	&	32			&	121117A	&	1	,	4	&	150910A	&	1	,	4	&	200829A	&	1	,	2	\\
060116	&	1	,	2	&	071117	&	30			&	091020	&	32			&	121128A	&	1	,	3	&	151021A	&	1	,	8	&	201020A	&	1	,	2	\\
060117	&	1	,	2	&	080205	&	1	,	3	&	091024	&	32			&	121209A	&	1	,	3	&	151027A	&	1	,	3	&	201021C	&	1	,	2	\\
060204B	&	1	,	3	&	080207	&	32			&	091029	&	32			&	121211A	&	1	,	3	&	151029A	&	1	,	4	&	201024A	&	1	,	2	\\
060210	&	32			&	080210	&	1	,	4	&	091109A	&	2	,	4	&	121217A	&	1	,	2	&	151031A	&	1	,	3	&	201216C	&	1	,	2	\\
060223A	&	2	,	3	&	080310	&	1	,	4	&	091127	&	32			&	130215A	&	1	,	3	&	151111A	&	1	,	4	&	201221A	&	1	,	2	\\
060306	&	32			&	080319A	&	1	,	3	&	091208B	&	1	,	3	&	130408A	&	1	,	3	&	151112A	&	1	,	4	&	210104A	&	1	,	2	\\
060319	&	1	,	4	&	080319B	&	30			&	100418A	&	1	,	4	&	130420A	&	33			&	151215A	&	1	,	3	&	210112A	&	1	,	2	\\
060418	&	30			&	080319C	&	30			&	100425A	&	1	,	4	&	130427A	&	33			&	160104A	&	1	,	2	&	210207B	&	1	,	2	\\
060501	&	3			&	080325	&	1	,	4	&	100606A	&	3			&	130427B	&	1	,	3	&	160121A	&	1	,	3	&	210210A	&	1	,	2	\\
060502A	&	3			&	080411	&	30			&	100615A	&	1	,	3	&	130505A	&	1	,	4	&	160131A	&	1	,	4	&	210217A	&	1	,	2	\\
060505	&	1			&	080413A	&	2	,	3	&	100621A	&	1	,	3	&	130528A	&	3			&	160203A	&	1	,	3	&	210321A	&	1	,	2	\\
060510A	&	3			&	080413B	&	31			&	100704A	&	2	,	3	&	130606A	&	1	,	3	&	160228A	&	1	,	3	&	210610A	&	1	,	2	\\
060526	&	30			&	080430	&	1	,	4	&	100728A	&	1	,	3	&	130610A	&	1	,	3	&	160327A	&	1	,	3	&	210610B	&	1	,	2	\\
060607A	&	2	,	3	&	080515	&	1	,	3	&	100728B	&	1	,	3	&	130701A	&	33			&	160425A	&	1			&	210619B	&	1	,	2	\\
060707	&	30			&	080602	&	1	,	4	&	100814A	&	1	,	3	&	130831A	&	33			&	160703A	&	1	,	2	&	210722A	&	1	,	2	\\
060708	&	1	,	3	&	080603B	&	31			&	100902A	&	2	,	4	&	130907A	&	33			&	160804A	&	1	,	9	&	210731A	&	1	,	2	\\
060714	&	1	,	4	&	080605	&	31			&	100906A	&	1	,	3	&	130925A	&	2	,	3	&	161014A	&	1	,	4	&	210822A	&	1	,	2	\\
060719	&	1	,	4	&	080607	&	31			&	110106B	&	1	,	3	&	131030A	&	33			&	161017A	&	1	,	10	&	211211A	&	1	,	2	\\
060729	&	30			&	080707	&	1	,	4	&	110205A	&	1	,	3	&	131103A	&	1	,	4	&	161117A	&	1	,	11	&	211227A	&	1	,	2	\\
060814	&	30			&	080721	&	31			&	110213A	&	1	,	3	&	131105A	&	33			&	161129A	&	1	,	4	&	220101A	&	1	,	2	\\
060904B	&	1	,	3	&	080804	&	32			&	110422A	&	1	,	3	&	131227A	&	1	,	4	&	170113A	&	1	,	12	&	220117A	&	1	,	2	\\
060906	&	1	,	4	&	080805	&	1	,	4	&	110503A	&	1	,	3	&	140206A	&	33	,	34	&	170202A	&	1	,	13	&	220521A	&	1	,	2	\\
060908	&	30			&	080810	&	31			&	110709B	&	1	,	3	&	140213A	&	33	,	34	&	170604A	&	1	,	4	&	230325A	&	1	,	2	\\
060912A	&	2	,	3	&	080906	&	1	,	3	&	110715A	&	1	,	3	&	140304A	&	1	,	3	&	170705A	&	1	,	14	&	230506C	&	1	,	2	\\
060923A	&	1	,	2	&	080916A	&	2	,	3	&	110726A	&	3			&	140311A	&	1	,	4	&	170903A	&	1			&	230818A	&	1	,	2	\\
060923B	&	1	,	4	&	080928	&	32			&	110731A	&	1	,	3	&	140419A	&	1	,	3	&	171209A	&	1	,	2	&	231111A	&	1	,	2	\\
060927	&	30			&	081007	&	31			&	110801A	&	1	,	3	&	140423A	&	1	,	3	&	180205A	&	1	,	15	&	231118A	&	1	,	2	\\
061007	&	30			&		&				&		&				&		&				&		&				&		&				\\

\end{longtable}
\noindent \textbf{Note:} 307 Swift GRB used in this study. The spectra and redshift references are from the following sources: 
(1) \url{https://swift.gsfc.nasa.gov/archive/grb_table/};
(2) \url{https://swift.gsfc.nasa.gov/archive/grbtable/};
 (3) \cite{2020ApJ...893...77W};
 (4) \cite{2007ApJ...671..656B};
 (5) \cite{2014GCN.16798....1Z};
 (6) \cite{2014GCN.17216....1Y};
 (7) \cite{2015GCN.17319....1V};
 (8) \cite{2015GCN.18433....1G};
 (9) \cite{2016GCN.19769....1B};
 (10) \cite{2016GCN.20082....1F};
 (11) \cite{2016GCN.20192....1M};
 (12) \cite{2017GCN.20456....1M};
 (13) \cite{2017GCN.20604....1F};
 (14) \cite{2017GCN.21297....1B};
 (15) \cite{2018GCN.22386....1V};
 (16) \cite{2018GCN.22513....1T};
 (17) \cite{2018GCN.22546....1F};
 (18) \cite{2018GCN.22566....1P};
 (19) \cite{2018GCN.22813....1P};
 (20) \cite{2018GCN.23053....1P};
 (21) \cite{2018GCN.23363....1T};
 (22) \cite{2018GCN.23424....1F};
 (23) \cite{2019GCN.23637....1T};
 (24) \cite{2019GCN.23707....1H};
 (25) \cite{2019GCN.24002....1H};
 (26) \cite{2019GCN.24816....1P};
 (27) \cite{2019GCN.25130....1P};
 (28) \cite{2019GCN.25974....1S};
 (29) \cite{2019GCN.26000....1B};
 (30) \cite{2008MNRAS.391..577A};
 (31) \cite{2009A&A...508..173A};
 (32) \cite{2019MNRAS.486L..46A};
 (33) \cite{2016A&A...585A..68W};
 (34) \cite{2020ApJ...893...46V}.






\bibliography{sample631}{}

\begin{thebibliography}{}
\expandafter\ifx\csname natexlab\endcsname\relax\def\natexlab#1{#1}\fi
\providecommand{\url}[1]{\href{#1}{#1}}
\providecommand{\dodoi}[1]{doi:~\href{http://doi.org/#1}{\nolinkurl{#1}}}
\providecommand{\doeprint}[1]{\href{http://ascl.net/#1}{\nolinkurl{http://ascl.net/#1}}}
\providecommand{\doarXiv}[1]{\href{https://arxiv.org/abs/#1}{\nolinkurl{https://arxiv.org/abs/#1}}}

\bibitem[{{ Mailyan}(2018)}]{2018GCN.22813....1P}
{ Mailyan}, B. 2018, GRB Coordinates Network, 22813, 1

\bibitem[{{Abbott} {et~al.}(2017){Abbott}, {Abbott}, {Abbott}, {Acernese}, {Ackley}, {Adams}, {Adams}, {Addesso}, {Adhikari}, {Adya}, {Affeldt}, {Afrough}, {Agarwal}, {Agathos}, {Agatsuma}, {Aggarwal}, {Aguiar}, {Aiello}, {Ain}, {Ajith}, {Allen}, {Allen}, {Allocca}, {Aloy}, {Altin}, {Amato}, {Ananyeva}, {Anderson}, {Anderson}, {Angelova}, {Antier}, {Appert}, {Arai}, {Araya}, {Areeda}, {Arnaud}, {Arun}, {Ascenzi}, {Ashton}, {Ast}, {Aston}, {Astone}, {Atallah}, {Aufmuth}, {Aulbert}, {AultONeal}, {Austin}, {Avila-Alvarez}, {Babak}, {Bacon}, {Bader}, {Bae}, {Baker}, {Baldaccini}, {Ballardin}, {Ballmer}, {Banagiri}, {Barayoga}, {Barclay}, {Barish}, {Barker}, {Barkett}, {Barone}, {Barr}, {Barsotti}, {Barsuglia}, {Barta}, {Bartlett}, {Bartos}, {Bassiri}, {Basti}, {Batch}, {Bawaj}, {Bayley}, {Bazzan}, {B{\'e}csy}, {Beer}, {Bejger}, {Belahcene}, {Bell}, {Berger}, {Bergmann}, {Bero}, {Berry}, {Bersanetti}, {Bertolini}, {Betzwieser}, {Bhagwat}, {Bhandare}, {Bilenko}, {Billingsley}, {Billman}, {Birch}, {Birney},
  {Birnholtz}, {Biscans}, {Biscoveanu}, {Bisht}, {Bitossi}, {Biwer}, {Bizouard}, {Blackburn}, {Blackman}, {Blair}, {Blair}, {Blair}, {Bloemen}, {Bock}, {Bode}, {Boer}, {Bogaert}, {Bohe}, {Bondu}, {Bonilla}, {Bonnand}, {Boom}, {Bork}, {Boschi}, {Bose}, {Bossie}, {Bouffanais}, {Bozzi}, {Bradaschia}, {Brady}, {Branchesi}, {Brau}, {Briant}, {Brillet}, {Brinkmann}, {Brisson}, {Brockill}, {Broida}, {Brooks}, {Brown}, {Brown}, {Brunett}, {Buchanan}, {Buikema}, {Bulik}, {Bulten}, {Buonanno}, {Buskulic}, {Buy}, {Byer}, {Cabero}, {Cadonati}, {Cagnoli}, {Cahillane}, {Calder{\'o}n Bustillo}, {Callister}, {Calloni}, {Camp}, {Canepa}, {Canizares}, {Cannon}, {Cao}, {Cao}, {Capano}, {Capocasa}, {Carbognani}, {Caride}, {Carney}, {Casanueva Diaz}, {Casentini}, {Caudill}, {Cavagli{\`a}}, {Cavalier}, {Cavalieri}, {Cella}, {Cepeda}, {Cerd{\'a}-Dur{\'a}n}, {Cerretani}, {Cesarini}, {Chamberlin}, {Chan}, {Chao}, {Charlton}, {Chase}, {Chassande-Mottin}, {Chatterjee}, {Chatziioannou}, {Cheeseboro}, {Chen}, {Chen}, {Chen}, {Cheng},
  {Chia}, {Chincarini}, {Chiummo}, {Chmiel}, {Cho}, {Cho}, {Chow}, {Christensen}, {Chu}, {Chua}, {Chua}, {Chung}, {Chung}, {Ciani}, {Ciolfi}, {Cirelli}, {Cirone}, {Clara}, {Clark}, {Clearwater}, {Cleva}, {Cocchieri}, {Coccia}, {Cohadon}, {Cohen}, {Colla}, {Collette}, {Cominsky}, {Constancio}, {Conti}, {Cooper}, {Corban}, {Corbitt}, {Cordero-Carri{\'o}n}, {Corley}, {Cornish}, {Corsi}, {Cortese}, {Costa}, {Coughlin}, {Coughlin}, {Coulon}, {Countryman}, {Couvares}, {Covas}, {Cowan}, {Coward}, {Cowart}, {Coyne}, {Coyne}, {Creighton}, {Creighton}, {Cripe}, {Crowder}, {Cullen}, {Cumming}, {Cunningham}, {Cuoco}, {Dal Canton}, {D{\'a}lya}, {Danilishin}, {D'Antonio}, {Danzmann}, {Dasgupta}, {Da Silva Costa}, {Dattilo}, {Dave}, {Davier}, {Davis}, {Daw}, {Day}, {De}, {DeBra}, {Degallaix}, {De Laurentis}, {Del{\'e}glise}, {Del Pozzo}, {Demos}, {Denker}, {Dent}, {De Pietri}, {Dergachev}, {De Rosa}, {DeRosa}, {De Rossi}, {DeSalvo}, {de Varona}, {Devenson}, {Dhurandhar}, {D{\'\i}az}, {Di Fiore}, {Di Giovanni}, {Di
  Girolamo}, {Di Lieto}, {Di Pace}, {Di Palma}, {Di Renzo}, {Doctor}, {Dolique}, {Donovan}, {Dooley}, {Doravari}, {Dorrington}, {Douglas}, {Dovale {\'A}lvarez}, {Downes}, {Drago}, {Dreissigacker}, {Driggers}, {Du}, {Ducrot}, {Dupej}, {Dwyer}, {Edo}, {Edwards}, {Effler}, {Eggenstein}, {Ehrens}, {Eichholz}, {Eikenberry}, {Eisenstein}, {Essick}, {Estevez}, {Etienne}, {Etzel}, {Evans}, {Evans}, {Factourovich}, {Fafone}, {Fair}, {Fairhurst}, {Fan}, {Farinon}, {Farr}, {Farr}, {Fauchon-Jones}, {Favata}, {Fays}, {Fee}, {Fehrmann}, {Feicht}, {Fejer}, {Fernandez-Galiana}, {Ferrante}, {Ferreira}, {Ferrini}, {Fidecaro}, {Finstad}, {Fiori}, {Fiorucci}, {Fishbach}, {Fisher}, {Fitz-Axen}, {Flaminio}, {Fletcher}, {Fong}, {Font}, {Forsyth}, {Forsyth}, {Fournier}, {Frasca}, {Frasconi}, {Frei}, {Freise}, {Frey}, {Frey}, {Fries}, {Fritschel}, {Frolov}, {Fulda}, {Fyffe}, {Gabbard}, {Gadre}, {Gaebel}, {Gair}, {Gammaitoni}, {Ganija}, {Gaonkar}, {Garcia-Quiros}, {Garufi}, {Gateley}, {Gaudio}, {Gaur}, {Gayathri}, {Gehrels}, {Gemme},
  {Genin}, {Gennai}, {George}, {George}, {Gergely}, {Germain}, {Ghonge}, {Ghosh}, {Ghosh}, {Ghosh}, {Giaime}, {Giardina}, {Giazotto}, {Gill}, {Glover}, {Goetz}, {Goetz}, {Gomes}, {Goncharov}, {Gonz{\'a}lez}, {Gonzalez Castro}, {Gopakumar}, {Gorodetsky}, {Gossan}, {Gosselin}, {Gouaty}, {Grado}, {Graef}, {Granata}, {Grant}, {Gras}, {Gray}, {Greco}, {Green}, {Gretarsson}, {Groot}, {Grote}, {Grunewald}, {Gruning}, {Guidi}, {Guo}, {Gupta}, {Gupta}, {Gushwa}, {Gustafson}, {Gustafson}, {Halim}, {Hall}, {Hall}, {Hamilton}, {Hammond}, {Haney}, {Hanke}, {Hanks}, {Hanna}, {Hannam}, {Hannuksela}, {Hanson}, {Hardwick}, {Harms}, {Harry}, {Harry}, {Hart}, {Haster}, {Haughian}, {Healy}, {Heidmann}, {Heintze}, {Heitmann}, {Hello}, {Hemming}, {Hendry}, {Heng}, {Hennig}, {Heptonstall}, {Heurs}, {Hild}, {Hinderer}, {Hoak}, {Hofman}, {Holt}, {Holz}, {Hopkins}, {Horst}, {Hough}, {Houston}, {Howell}, {Hreibi}, {Hu}, {Huerta}, {Huet}, {Hughey}, {Husa}, {Huttner}, {Huynh-Dinh}, {Indik}, {Inta}, {Intini}, {Isa}, {Isac}, {Isi}, {Iyer},
  {Izumi}, {Jacqmin}, {Jani}, {Jaranowski}, {Jawahar}, {Jim{\'e}nez-Forteza}, {Johnson}, {Johnson-McDaniel}, {Jones}, {Jones}, {Jonker}, {Ju}, {Junker}, {Kalaghatgi}, {Kalogera}, {Kamai}, {Kandhasamy}, {Kang}, {Kanner}, {Kapadia}, {Karki}, {Karvinen}, {Kasprzack}, {Kastaun}, {Katolik}, {Katsavounidis}, {Katzman}, {Kaufer}, {Kawabe}, {K{\'e}f{\'e}lian}, {Keitel}, {Kemball}, {Kennedy}, {Kent}, {Key}, {Khalili}, {Khan}, {Khan}, {Khan}, {Khazanov}, {Kijbunchoo}, {Kim}, {Kim}, {Kim}, {Kim}, {Kim}, {Kim}, {Kimbrell}, {King}, {King}, {Kinley-Hanlon}, {Kirchhoff}, {Kissel}, {Kleybolte}, {Klimenko}, {Knowles}, {Koch}, {Koehlenbeck}, {Koley}, {Kondrashov}, {Kontos}, {Korobko}, {Korth}, {Kowalska}, {Kozak}, {Kr{\"a}mer}, {Kringel}, {Krishnan}, {Kr{\'o}lak}, {Kuehn}, {Kumar}, {Kumar}, {Kumar}, {Kuo}, {Kutynia}, {Kwang}, {Lackey}, {Lai}, {Landry}, {Lang}, {Lange}, {Lantz}, {Lanza}, {Lartaux-Vollard}, {Lasky}, {Laxen}, {Lazzarini}, {Lazzaro}, {Leaci}, {Leavey}, {Lee}, {Lee}, {Lee}, {Lee}, {Lee}, {Lehmann}, {Lenon},
  {Leonardi}, {Leroy}, {Letendre}, {Levin}, {Li}, {Linker}, {Littenberg}, {Liu}, {Lo}, {Lockerbie}, {London}, {Lord}, {Lorenzini}, {Loriette}, {Lormand}, {Losurdo}, {Lough}, {Lousto}, {Lovelace}, {L{\"u}ck}, {Lumaca}, {Lundgren}, {Lynch}, {Ma}, {Macas}, {Macfoy}, {Machenschalk}, {MacInnis}, {Macleod}, {Maga{\~n}a Hernandez}, {Maga{\~n}a-Sandoval}, {Maga{\~n}a Zertuche}, {Magee}, {Majorana}, {Maksimovic}, {Man}, {Mandic}, {Mangano}, {Mansell}, {Manske}, {Mantovani}, {Marchesoni}, {Marion}, {M{\'a}rka}, {M{\'a}rka}, {Markakis}, {Markosyan}, {Markowitz}, {Maros}, {Marquina}, {Martelli}, {Martellini}, {Martin}, {Martin}, {Martynov}, {Mason}, {Massera}, {Masserot}, {Massinger}, {Masso-Reid}, {Mastrogiovanni}, {Matas}, {Matichard}, {Matone}, {Mavalvala}, {Mazumder}, {McCarthy}, {McClelland}, {McCormick}, {McCuller}, {McGuire}, {McIntyre}, {McIver}, {McManus}, {McNeill}, {McRae}, {McWilliams}, {Meacher}, {Meadors}, {Mehmet}, {Meidam}, {Mejuto-Villa}, {Melatos}, {Mendell}, {Mercer}, {Merilh}, {Merzougui}, {Meshkov},
  {Messenger}, {Messick}, {Metzdorff}, {Meyers}, {Miao}, {Michel}, {Middleton}, {Mikhailov}, {Milano}, {Miller}, {Miller}, {Miller}, {Millhouse}, {Milovich-Goff}, {Minazzoli}, {Minenkov}, {Ming}, {Mishra}, {Mitra}, {Mitrofanov}, {Mitselmakher}, {Mittleman}, {Moffa}, {Moggi}, {Mogushi}, {Mohan}, {Mohapatra}, {Montani}, {Moore}, {Moraru}, {Moreno}, {Morriss}, {Mours}, {Mow-Lowry}, {Mueller}, {Muir}, {Mukherjee}, {Mukherjee}, {Mukherjee}, {Mukund}, {Mullavey}, {Munch}, {Mu{\~n}iz}, {Muratore}, {Murray}, {Napier}, {Nardecchia}, {Naticchioni}, {Nayak}, {Neilson}, {Nelemans}, {Nelson}, {Nery}, {Neunzert}, {Nevin}, {Newport}, {Newton}, {Ng}, {Nguyen}, {Nichols}, {Nielsen}, {Nissanke}, {Nitz}, {Noack}, {Nocera}, {Nolting}, {North}, {Nuttall}, {Oberling}, {O'Dea}, {Ogin}, {Oh}, {Oh}, {Ohme}, {Okada}, {Oliver}, {Oppermann}, {Oram}, {O'Reilly}, {Ormiston}, {Ortega}, {O'Shaughnessy}, {Ossokine}, {Ottaway}, {Overmier}, {Owen}, {Pace}, {Page}, {Page}, {Pai}, {Pai}, {Palamos}, {Palashov}, {Palomba}, {Pal-Singh}, {Pan},
  {Pan}, {Pang}, {Pang}, {Pankow}, {Pannarale}, {Pant}, {Paoletti}, {Paoli}, {Papa}, {Parida}, {Parker}, {Pascucci}, {Pasqualetti}, {Passaquieti}, {Passuello}, {Patil}, {Patricelli}, {Pearlstone}, {Pedraza}, {Pedurand}, {Pekowsky}, {Pele}, {Penn}, {Perez}, {Perreca}, {Perri}, {Pfeiffer}, {Phelps}, {Piccinni}, {Pichot}, {Piergiovanni}, {Pierro}, {Pillant}, {Pinard}, {Pinto}, {Pirello}, {Pitkin}, {Poe}, {Poggiani}, {Popolizio}, {Porter}, {Post}, {Powell}, {Prasad}, {Pratt}, {Pratten}, {Predoi}, {Prestegard}, {Prijatelj}, {Principe}, {Privitera}, {Prodi}, {Prokhorov}, {Puncken}, {Punturo}, {Puppo}, {P{\"u}rrer}, {Qi}, {Quetschke}, {Quintero}, {Quitzow-James}, {Raab}, {Rabeling}, {Radkins}, {Raffai}, {Raja}, {Rajan}, {Rajbhandari}, {Rakhmanov}, {Ramirez}, {Ramos-Buades}, {Rapagnani}, {Raymond}, {Razzano}, {Read}, {Regimbau}, {Rei}, {Reid}, {Reitze}, {Ren}, {Reyes}, {Ricci}, {Ricker}, {Rieger}, {Riles}, {Rizzo}, {Robertson}, {Robie}, {Robinet}, {Rocchi}, {Rolland}, {Rollins}, {Roma}, {Romano}, {Romel}, {Romie},
  {Rosi{\'n}ska}, {Ross}, {Rowan}, {R{\"u}diger}, {Ruggi}, {Rutins}, {Ryan}, {Sachdev}, {Sadecki}, {Sadeghian}, {Sakellariadou}, {Salconi}, {Saleem}, {Salemi}, {Samajdar}, {Sammut}, {Sampson}, {Sanchez}, {Sanchez}, {Sanchis-Gual}, {Sandberg}, {Sanders}, {Sassolas}, {Sathyaprakash}, {Saulson}, {Sauter}, {Savage}, {Sawadsky}, {Schale}, {Scheel}, {Scheuer}, {Schmidt}, {Schmidt}, {Schnabel}, {Schofield}, {Sch{\"o}nbeck}, {Schreiber}, {Schuette}, {Schulte}, {Schutz}, {Schwalbe}, {Scott}, {Scott}, {Seidel}, {Sellers}, {Sengupta}, {Sentenac}, {Sequino}, {Sergeev}, {Shaddock}, {Shaffer}, {Shah}, {Shahriar}, {Shaner}, {Shao}, {Shapiro}, {Shawhan}, {Sheperd}, {Shoemaker}, {Shoemaker}, {Siellez}, {Siemens}, {Sieniawska}, {Sigg}, {Silva}, {Singer}, {Singh}, {Singhal}, {Sintes}, {Slagmolen}, {Smith}, {Smith}, {Smith}, {Somala}, {Son}, {Sonnenberg}, {Sorazu}, {Sorrentino}, {Souradeep}, {Spencer}, {Srivastava}, {Staats}, {Staley}, {Steinke}, {Steinlechner}, {Steinlechner}, {Steinmeyer}, {Stevenson}, {Stone}, {Stops},
  {Strain}, {Stratta}, {Strigin}, {Strunk}, {Sturani}, {Stuver}, {Summerscales}, {Sun}, {Sunil}, {Suresh}, {Sutton}, {Swinkels}, {Szczepa{\'n}czyk}, {Tacca}, {Tait}, {Talbot}, {Talukder}, {Tanner}, {T{\'a}pai}, {Taracchini}, {Tasson}, {Taylor}, {Taylor}, {Tewari}, {Theeg}, {Thies}, {Thomas}, {Thomas}, {Thomas}, {Thorne}, {Thorne}, {Thrane}, {Tiwari}, {Tiwari}, {Tokmakov}, {Toland}, {Tonelli}, {Tornasi}, {Torres-Forn{\'e}}, {Torrie}, {T{\"o}yr{\"a}}, {Travasso}, {Traylor}, {Trinastic}, {Tringali}, {Trozzo}, {Tsang}, {Tse}, {Tso}, {Tsukada}, {Tsuna}, {Tuyenbayev}, {Ueno}, {Ugolini}, {Unnikrishnan}, {Urban}, {Usman}, {Vahlbruch}, {Vajente}, {Valdes}, {van Bakel}, {van Beuzekom}, {van den Brand}, {Van Den Broeck}, {Vander-Hyde}, {van der Schaaf}, {van Heijningen}, {van Veggel}, {Vardaro}, {Varma}, {Vass}, {Vas{\'u}th}, {Vecchio}, {Vedovato}, {Veitch}, {Veitch}, {Venkateswara}, {Venugopalan}, {Verkindt}, {Vetrano}, {Vicer{\'e}}, {Viets}, {Vinciguerra}, {Vine}, {Vinet}, {Vitale}, {Vo}, {Vocca}, {Vorvick},
  {Vyatchanin}, {Wade}, {Wade}, {Wade}, {Walet}, {Walker}, {Wallace}, {Walsh}, {Wang}, {Wang}, {Wang}, {Wang}, {Wang}, {Ward}, {Warner}, {Was}, {Watchi}, {Weaver}, {Wei}, {Weinert}, {Weinstein}, {Weiss}, {Wen}, {Wessel}, {We{\ss}els}, {Westerweck}, {Westphal}, {Wette}, {Whelan}, {Whitcomb}, {Whiting}, {Whittle}, {Wilken}, {Williams}, {Williams}, {Williamson}, {Willis}, {Willke}, {Wimmer}, {Winkler}, {Wipf}, {Wittel}, {Woan}, {Woehler}, {Wofford}, {Wong}, {Worden}, {Wright}, {Wu}, {Wysocki}, {Xiao}, {Yamamoto}, {Yancey}, {Yang}, {Yap}, {Yazback}, {Yu}, {Yu}, {Yvert}, {Zadro{\.z}ny}, {Zanolin}, {Zelenova}, {Zendri}, {Zevin}, {Zhang}, {Zhang}, {Zhang}, {Zhang}, {Zhao}, {Zhou}, {Zhou}, {Zhu}, {Zhu}, {Zimmerman}, {Zucker}, {Zweizig}, {(LIGO Scientific Collaboration}, {Virgo Collaboration}, {Burns}, {Veres}, {Kocevski}, {Racusin}, {Goldstein}, {Connaughton}, {Briggs}, {Blackburn}, {Hamburg}, {Hui}, {von Kienlin}, {McEnery}, {Preece}, {Wilson-Hodge}, {Bissaldi}, {Cleveland}, {Gibby}, {Giles}, {Kippen}, {McBreen},
  {Meegan}, {Paciesas}, {Poolakkil}, {Roberts}, {Stanbro}, {Gamma-ray Burst Monitor}, {Savchenko}, {Ferrigno}, {Kuulkers}, {Bazzano}, {Bozzo}, {Brandt}, {Chenevez}, {Courvoisier}, {Diehl}, {Domingo}, {Hanlon}, {Jourdain}, {Laurent}, {Lebrun}, {Lutovinov}, {Mereghetti}, {Natalucci}, {Rodi}, {Roques}, {Sunyaev}, {Ubertini}, \& {(INTEGRAL}}]{2017ApJ...848L..13A}
{Abbott}, B.~P., {Abbott}, R., {Abbott}, T.~D., {et~al.} 2017, \apjl, 848, L13, \dodoi{10.3847/2041-8213/aa920c}

\bibitem[{{Abdo} {et~al.}(2010){Abdo}, {Ackermann}, {Ajello}, {Antolini}, {Baldini}, {Ballet}, {Barbiellini}, {Bastieri}, {Baughman}, {Bechtol}, {Bellazzini}, {Berenji}, {Blandford}, {Bloom}, {Bonamente}, {Borgland}, {Bouvier}, {Bregeon}, {Brez}, {Brigida}, {Bruel}, {Burnett}, {Buson}, {Caliandro}, {Cameron}, {Caraveo}, {Carrigan}, {Casandjian}, {Cavazzuti}, {Cecchi}, {{\c{C}}elik}, {Charles}, {Chekhtman}, {Cheung}, {Chiang}, {Ciprini}, {Claus}, {Cohen-Tanugi}, {Conrad}, {Costamante}, {Cutini}, {Dermer}, {de Angelis}, {de Palma}, {Silva}, {Drell}, {Dubois}, {Dumora}, {Farnier}, {Favuzzi}, {Fegan}, {Focke}, {Fukazawa}, {Funk}, {Fusco}, {Gargano}, {Gasparrini}, {Gehrels}, {Germani}, {Giglietto}, {Giommi}, {Giordano}, {Glanzman}, {Godfrey}, {Grenier}, {Grove}, {Guiriec}, {Hadasch}, {Hayashida}, {Hays}, {Healey}, {Horan}, {Hughes}, {Itoh}, {J{\'o}hannesson}, {Johnson}, {Johnson}, {Johnson}, {Kamae}, {Katagiri}, {Kataoka}, {Kawai}, {Kn{\"o}dlseder}, {Kuss}, {Lande}, {Latronico}, {Lee}, {Lemoine-Goumard}, {Llena
  Garde}, {Longo}, {Loparco}, {Lott}, {Lovellette}, {Lubrano}, {Madejski}, {Makeev}, {Mazziotta}, {McConville}, {McEnery}, {Meurer}, {Michelson}, {Mitthumsiri}, {Mizuno}, {Monte}, {Monzani}, {Morselli}, {Moskalenko}, {Murgia}, {Nolan}, {Norris}, {Nuss}, {Ohsugi}, {Omodei}, {Orlando}, {Ormes}, {Ozaki}, {Paneque}, {Panetta}, {Parent}, {Pelassa}, {Pepe}, {Pesce-Rollins}, {Piron}, {Porter}, {Rain{\`o}}, {Rando}, {Razzano}, {Reimer}, {Reimer}, {Ritz}, {Rochester}, {Rodriguez}, {Romani}, {Roth}, {Sadrozinski}, {Sander}, {Saz Parkinson}, {Scargle}, {Sgr{\`o}}, {Shaw}, {Smith}, {Spandre}, {Spinelli}, {Starck}, {Strickman}, {Strong}, {Suson}, {Tajima}, {Takahashi}, {Takahashi}, {Tanaka}, {Thayer}, {Thayer}, {Thompson}, {Tibaldo}, {Torres}, {Tosti}, {Tramacere}, {Uchiyama}, {Usher}, {Vasileiou}, {Vilchez}, {Vitale}, {Waite}, {Wang}, {Winer}, {Wood}, {Yang}, {Ylinen}, {Ziegler}, \& {Fermi LAT Collaboration}}]{2010ApJ...720..435A}
{Abdo}, A.~A., {Ackermann}, M., {Ajello}, M., {et~al.} 2010, \apj, 720, 435, \dodoi{10.1088/0004-637X/720/1/435}

\bibitem[{{Ajello} {et~al.}(2009){Ajello}, {Costamante}, {Sambruna}, {Gehrels}, {Chiang}, {Rau}, {Escala}, {Greiner}, {Tueller}, {Wall}, \& {Mushotzky}}]{2009ApJ...699..603A}
{Ajello}, M., {Costamante}, L., {Sambruna}, R.~M., {et~al.} 2009, \apj, 699, 603, \dodoi{10.1088/0004-637X/699/1/603}

\bibitem[{{Ajello} {et~al.}(2012){Ajello}, {Shaw}, {Romani}, {Dermer}, {Costamante}, {King}, {Max-Moerbeck}, {Readhead}, {Reimer}, {Richards}, \& {Stevenson}}]{2012ApJ...751..108A}
{Ajello}, M., {Shaw}, M.~S., {Romani}, R.~W., {et~al.} 2012, \apj, 751, 108, \dodoi{10.1088/0004-637X/751/2/108}

\bibitem[{{Akaike}(1974)}]{1974ITAC...19..716A}
{Akaike}, H. 1974, IEEE Transactions on Automatic Control, 19, 716

\bibitem[{{Amati}(2006)}]{2006MNRAS.372..233A}
{Amati}, L. 2006, \mnras, 372, 233, \dodoi{10.1111/j.1365-2966.2006.10840.x}

\bibitem[{{Amati} {et~al.}(2019){Amati}, {D'Agostino}, {Luongo}, {Muccino}, \& {Tantalo}}]{2019MNRAS.486L..46A}
{Amati}, L., {D'Agostino}, R., {Luongo}, O., {Muccino}, M., \& {Tantalo}, M. 2019, \mnras, 486, L46, \dodoi{10.1093/mnrasl/slz056}

\bibitem[{{Amati} {et~al.}(2009){Amati}, {Frontera}, \& {Guidorzi}}]{2009A&A...508..173A}
{Amati}, L., {Frontera}, F., \& {Guidorzi}, C. 2009, \aap, 508, 173, \dodoi{10.1051/0004-6361/200912788}

\bibitem[{{Amati} {et~al.}(2008){Amati}, {Guidorzi}, {Frontera}, {Della Valle}, {Finelli}, {Landi}, \& {Montanari}}]{2008MNRAS.391..577A}
{Amati}, L., {Guidorzi}, C., {Frontera}, F., {et~al.} 2008, \mnras, 391, 577, \dodoi{10.1111/j.1365-2966.2008.13943.x}

\bibitem[{{Amati} {et~al.}(2002){Amati}, {Frontera}, {Tavani}, {in't Zand}, {Antonelli}, {Costa}, {Feroci}, {Guidorzi}, {Heise}, {Masetti}, {Montanari}, {Nicastro}, {Palazzi}, {Pian}, {Piro}, \& {Soffitta}}]{2002A&A...390...81A}
{Amati}, L., {Frontera}, F., {Tavani}, M., {et~al.} 2002, \aap, 390, 81, \dodoi{10.1051/0004-6361:20020722}

\bibitem[{{Astropy Collaboration} {et~al.}(2013){Astropy Collaboration}, {Robitaille}, {Tollerud}, {Greenfield}, {Droettboom}, {Bray}, {Aldcroft}, {Davis}, {Ginsburg}, {Price-Whelan}, {Kerzendorf}, {Conley}, {Crighton}, {Barbary}, {Muna}, {Ferguson}, {Grollier}, {Parikh}, {Nair}, {Unther}, {Deil}, {Woillez}, {Conseil}, {Kramer}, {Turner}, {Singer}, {Fox}, {Weaver}, {Zabalza}, {Edwards}, {Azalee Bostroem}, {Burke}, {Casey}, {Crawford}, {Dencheva}, {Ely}, {Jenness}, {Labrie}, {Lim}, {Pierfederici}, {Pontzen}, {Ptak}, {Refsdal}, {Servillat}, \& {Streicher}}]{2013A&A...558A..33A}
{Astropy Collaboration}, {Robitaille}, T.~P., {Tollerud}, E.~J., {et~al.} 2013, \aap, 558, A33, \dodoi{10.1051/0004-6361/201322068}

\bibitem[{{Astropy Collaboration} {et~al.}(2018){Astropy Collaboration}, {Price-Whelan}, {Sip{\H{o}}cz}, {G{\"u}nther}, {Lim}, {Crawford}, {Conseil}, {Shupe}, {Craig}, {Dencheva}, {Ginsburg}, {VanderPlas}, {Bradley}, {P{\'e}rez-Su{\'a}rez}, {de Val-Borro}, {Aldcroft}, {Cruz}, {Robitaille}, {Tollerud}, {Ardelean}, {Babej}, {Bach}, {Bachetti}, {Bakanov}, {Bamford}, {Barentsen}, {Barmby}, {Baumbach}, {Berry}, {Biscani}, {Boquien}, {Bostroem}, {Bouma}, {Brammer}, {Bray}, {Breytenbach}, {Buddelmeijer}, {Burke}, {Calderone}, {Cano Rodr{\'\i}guez}, {Cara}, {Cardoso}, {Cheedella}, {Copin}, {Corrales}, {Crichton}, {D'Avella}, {Deil}, {Depagne}, {Dietrich}, {Donath}, {Droettboom}, {Earl}, {Erben}, {Fabbro}, {Ferreira}, {Finethy}, {Fox}, {Garrison}, {Gibbons}, {Goldstein}, {Gommers}, {Greco}, {Greenfield}, {Groener}, {Grollier}, {Hagen}, {Hirst}, {Homeier}, {Horton}, {Hosseinzadeh}, {Hu}, {Hunkeler}, {Ivezi{\'c}}, {Jain}, {Jenness}, {Kanarek}, {Kendrew}, {Kern}, {Kerzendorf}, {Khvalko}, {King}, {Kirkby}, {Kulkarni},
  {Kumar}, {Lee}, {Lenz}, {Littlefair}, {Ma}, {Macleod}, {Mastropietro}, {McCully}, {Montagnac}, {Morris}, {Mueller}, {Mumford}, {Muna}, {Murphy}, {Nelson}, {Nguyen}, {Ninan}, {N{\"o}the}, {Ogaz}, {Oh}, {Parejko}, {Parley}, {Pascual}, {Patil}, {Patil}, {Plunkett}, {Prochaska}, {Rastogi}, {Reddy Janga}, {Sabater}, {Sakurikar}, {Seifert}, {Sherbert}, {Sherwood-Taylor}, {Shih}, {Sick}, {Silbiger}, {Singanamalla}, {Singer}, {Sladen}, {Sooley}, {Sornarajah}, {Streicher}, {Teuben}, {Thomas}, {Tremblay}, {Turner}, {Terr{\'o}n}, {van Kerkwijk}, {de la Vega}, {Watkins}, {Weaver}, {Whitmore}, {Woillez}, {Zabalza}, \& {Astropy Contributors}}]{2018AJ....156..123A}
{Astropy Collaboration}, {Price-Whelan}, A.~M., {Sip{\H{o}}cz}, B.~M., {et~al.} 2018, \aj, 156, 123, \dodoi{10.3847/1538-3881/aabc4f}

\bibitem[{{Astropy Collaboration} {et~al.}(2022){Astropy Collaboration}, {Price-Whelan}, {Lim}, {Earl}, {Starkman}, {Bradley}, {Shupe}, {Patil}, {Corrales}, {Brasseur}, {N{"o}the}, {Donath}, {Tollerud}, {Morris}, {Ginsburg}, {Vaher}, {Weaver}, {Tocknell}, {Jamieson}, {van Kerkwijk}, {Robitaille}, {Merry}, {Bachetti}, {G{"u}nther}, {Aldcroft}, {Alvarado-Montes}, {Archibald}, {B{'o}di}, {Bapat}, {Barentsen}, {Baz{'a}n}, {Biswas}, {Boquien}, {Burke}, {Cara}, {Cara}, {Conroy}, {Conseil}, {Craig}, {Cross}, {Cruz}, {D'Eugenio}, {Dencheva}, {Devillepoix}, {Dietrich}, {Eigenbrot}, {Erben}, {Ferreira}, {Foreman-Mackey}, {Fox}, {Freij}, {Garg}, {Geda}, {Glattly}, {Gondhalekar}, {Gordon}, {Grant}, {Greenfield}, {Groener}, {Guest}, {Gurovich}, {Handberg}, {Hart}, {Hatfield-Dodds}, {Homeier}, {Hosseinzadeh}, {Jenness}, {Jones}, {Joseph}, {Kalmbach}, {Karamehmetoglu}, {Ka{l}uszy{'n}ski}, {Kelley}, {Kern}, {Kerzendorf}, {Koch}, {Kulumani}, {Lee}, {Ly}, {Ma}, {MacBride}, {Maljaars}, {Muna}, {Murphy}, {Norman}, {O'Steen},
  {Oman}, {Pacifici}, {Pascual}, {Pascual-Granado}, {Patil}, {Perren}, {Pickering}, {Rastogi}, {Roulston}, {Ryan}, {Rykoff}, {Sabater}, {Sakurikar}, {Salgado}, {Sanghi}, {Saunders}, {Savchenko}, {Schwardt}, {Seifert-Eckert}, {Shih}, {Jain}, {Shukla}, {Sick}, {Simpson}, {Singanamalla}, {Singer}, {Singhal}, {Sinha}, {Sip{H{o}}cz}, {Spitler}, {Stansby}, {Streicher}, {{{S}}umak}, {Swinbank}, {Taranu}, {Tewary}, {Tremblay}, {Val-Borro}, {Van Kooten}, {Vasovi{'c}}, {Verma}, {de Miranda Cardoso}, {Williams}, {Wilson}, {Winkel}, {Wood-Vasey}, {Xue}, {Yoachim}, {Zhang}, {Zonca}, \& {Astropy Project Contributors}}]{astropy:2022}
{Astropy Collaboration}, {Price-Whelan}, A.~M., {Lim}, P.~L., {et~al.} 2022, \apj, 935, 167, \dodoi{10.3847/1538-4357/ac7c74}

\bibitem[{{Band} {et~al.}(1993){Band}, {Matteson}, {Ford}, {Schaefer}, {Palmer}, {Teegarden}, {Cline}, {Briggs}, {Paciesas}, {Pendleton}, {Fishman}, {Kouveliotou}, {Meegan}, {Wilson}, \& {Lestrade}}]{1993ApJ...413..281B}
{Band}, D., {Matteson}, J., {Ford}, L., {et~al.} 1993, \apj, 413, 281, \dodoi{10.1086/172995}

\bibitem[{{Band}(2006)}]{2006ApJ...644..378B}
{Band}, D.~L. 2006, \apj, 644, 378, \dodoi{10.1086/503326}

\bibitem[{{Barthelmy} {et~al.}(2005){Barthelmy}, {Barbier}, {Cummings}, {Fenimore}, {Gehrels}, {Hullinger}, {Krimm}, {Markwardt}, {Palmer}, {Parsons}, {Sato}, {Suzuki}, {Takahashi}, {Tashiro}, \& {Tueller}}]{2005SSRv..120..143B}
{Barthelmy}, S.~D., {Barbier}, L.~M., {Cummings}, J.~R., {et~al.} 2005, \ssr, 120, 143, \dodoi{10.1007/s11214-005-5096-3}

\bibitem[{{Berger}(2014)}]{2014ARA&A..52...43B}
{Berger}, E. 2014, \araa, 52, 43, \dodoi{10.1146/annurev-astro-081913-035926}

\bibitem[{{Bissaldi} {et~al.}(2016){Bissaldi}, {Hui}, {Connaughton}, \& {Hamburg}}]{2016GCN.19769....1B}
{Bissaldi}, E., {Hui}, C.~M., {Connaughton}, V., \& {Hamburg}, R. 2016, GRB Coordinates Network, 19769, 1

\bibitem[{{Bissaldi} \& {Meegan}(2017)}]{2017GCN.21297....1B}
{Bissaldi}, E., \& {Meegan}, C. 2017, GRB Coordinates Network, 21297, 1

\bibitem[{{Bissaldi} {et~al.}(2019){Bissaldi}, {Veres}, \& {Fermi GBM Team}}]{2019GCN.26000....1B}
{Bissaldi}, E., {Veres}, P., \& {Fermi GBM Team}. 2019, GRB Coordinates Network, 26000, 1

\bibitem[{{Bloom}(2003)}]{2003AJ....125.2865B}
{Bloom}, J.~S. 2003, \aj, 125, 2865, \dodoi{10.1086/374945}

\bibitem[{{Bloom} {et~al.}(1998){Bloom}, {Djorgovski}, {Kulkarni}, \& {Frail}}]{1998ApJ...507L..25B}
{Bloom}, J.~S., {Djorgovski}, S.~G., {Kulkarni}, S.~R., \& {Frail}, D.~A. 1998, \apjl, 507, L25, \dodoi{10.1086/311682}

\bibitem[{{Bromberg} {et~al.}(2012){Bromberg}, {Nakar}, {Piran}, \& {Sari}}]{2012ApJ...749..110B}
{Bromberg}, O., {Nakar}, E., {Piran}, T., \& {Sari}, R. 2012, \apj, 749, 110, \dodoi{10.1088/0004-637X/749/2/110}

\bibitem[{{Bromberg} {et~al.}(2013){Bromberg}, {Nakar}, {Piran}, \& {Sari}}]{2013ApJ...764..179B}
---. 2013, \apj, 764, 179, \dodoi{10.1088/0004-637X/764/2/179}

\bibitem[{{Butler} {et~al.}(2007){Butler}, {Kocevski}, {Bloom}, \& {Curtis}}]{2007ApJ...671..656B}
{Butler}, N.~R., {Kocevski}, D., {Bloom}, J.~S., \& {Curtis}, J.~L. 2007, \apj, 671, 656, \dodoi{10.1086/522492}

\bibitem[{{Cano} {et~al.}(2017){Cano}, {Wang}, {Dai}, \& {Wu}}]{2017AdAst2017E...5C}
{Cano}, Z., {Wang}, S.-Q., {Dai}, Z.-G., \& {Wu}, X.-F. 2017, Advances in Astronomy, 2017, 8929054, \dodoi{10.1155/2017/8929054}

\bibitem[{{Cucchiara} {et~al.}(2011){Cucchiara}, {Levan}, {Fox}, {Tanvir}, {Ukwatta}, {Berger}, {Kr{\"u}hler}, {K{\"u}pc{\"u} Yolda{\c{s}}}, {Wu}, {Toma}, {Greiner}, {Olivares}, {Rowlinson}, {Amati}, {Sakamoto}, {Roth}, {Stephens}, {Fritz}, {Fynbo}, {Hjorth}, {Malesani}, {Jakobsson}, {Wiersema}, {O'Brien}, {Soderberg}, {Foley}, {Fruchter}, {Rhoads}, {Rutledge}, {Schmidt}, {Dopita}, {Podsiadlowski}, {Willingale}, {Wolf}, {Kulkarni}, \& {D'Avanzo}}]{2011ApJ7367C}
{Cucchiara}, A., {Levan}, A.~J., {Fox}, D.~B., {et~al.} 2011, \apj, 736, 7, \dodoi{10.1088/0004-637X/736/1/7}

\bibitem[{{Dainotti} {et~al.}(2022){Dainotti}, {De Simone}, {Islam}, {Kawaguchi}, {Moriya}, {Takiwaki}, {Tominaga}, \& {Gangopadhyay}}]{2022ApJ...938...41D}
{Dainotti}, M.~G., {De Simone}, B., {Islam}, K.~M., {et~al.} 2022, \apj, 938, 41, \dodoi{10.3847/1538-4357/ac8b77}

\bibitem[{{Dainotti} {et~al.}(2021){Dainotti}, {Petrosian}, \& {Bowden}}]{2021ApJ...914L..40D}
{Dainotti}, M.~G., {Petrosian}, V., \& {Bowden}, L. 2021, \apjl, 914, L40, \dodoi{10.3847/2041-8213/abf5e4}

\bibitem[{{Dong} {et~al.}(2022){Dong}, {Li}, {Zhang}, \& {Zhang}}]{2022MNRAS.513.1078D}
{Dong}, X.~F., {Li}, X.~J., {Zhang}, Z.~B., \& {Zhang}, X.~L. 2022, \mnras, 513, 1078, \dodoi{10.1093/mnras/stac949}

\bibitem[{{Dong} {et~al.}(2023){Dong}, {Zhang}, {Li}, {Huang}, \& {Bian}}]{2023ApJ...958...37D}
{Dong}, X.~F., {Zhang}, Z.~B., {Li}, Q.~M., {Huang}, Y.~F., \& {Bian}, K. 2023, \apj, 958, 37, \dodoi{10.3847/1538-4357/acf852}

\bibitem[{{Elliott} {et~al.}(2012){Elliott}, {Greiner}, {Khochfar}, {Schady}, {Johnson}, \& {Rau}}]{2012A&A...539A.113E}
{Elliott}, J., {Greiner}, J., {Khochfar}, S., {et~al.} 2012, \aap, 539, A113, \dodoi{10.1051/0004-6361/201118561}

\bibitem[{{Foreman-Mackey} {et~al.}(2013){Foreman-Mackey}, {Hogg}, {Lang}, \& {Goodman}}]{2013PASP..125..306F}
{Foreman-Mackey}, D., {Hogg}, D.~W., {Lang}, D., \& {Goodman}, J. 2013, \pasp, 125, 306, \dodoi{10.1086/670067}

\bibitem[{{Frederiks} {et~al.}(2016){Frederiks}, {Golenetskii}, {Aptekar}, {Oleynik}, {Ulanov}, {Svinkin}, {Tsvetkova}, {Lysenko}, {Kozlova}, \& {Cline}}]{2016GCN.20082....1F}
{Frederiks}, D., {Golenetskii}, S., {Aptekar}, R., {et~al.} 2016, GRB Coordinates Network, 20082, 1

\bibitem[{{Frederiks} {et~al.}(2017){Frederiks}, {Golenetskii}, {Aptekar}, {Oleynik}, {Ulanov}, {Svinkin}, {Tsvetkova}, {Lysenko}, {Kozlova}, \& {Cline}}]{2017GCN.20604....1F}
---. 2017, GRB Coordinates Network, 20604, 1

\bibitem[{{Frederiks} {et~al.}(2018{\natexlab{a}}){Frederiks}, {Golenetskii}, {Aptekar}, {Ulanov}, {Svinkin}, {Tsvetkova}, {Lysenko}, {Kozlova}, \& {Cline}}]{2018GCN.22546....1F}
---. 2018{\natexlab{a}}, GRB Coordinates Network, 22546, 1

\bibitem[{{Frederiks} {et~al.}(2018{\natexlab{b}}){Frederiks}, {Golenetskii}, {Aptekar}, {Kozlova}, {Lysenko}, {Svinkin}, {Tsvetkova}, {Ulanov}, \& {Cline}}]{2018GCN.23424....1F}
---. 2018{\natexlab{b}}, GRB Coordinates Network, 23424, 1

\bibitem[{{Galama} {et~al.}(1998){Galama}, {Vreeswijk}, {van Paradijs}, {Kouveliotou}, {Augusteijn}, {B{\"o}hnhardt}, {Brewer}, {Doublier}, {Gonzalez}, {Leibundgut}, {Lidman}, {Hainaut}, {Patat}, {Heise}, {in't Zand}, {Hurley}, {Groot}, {Strom}, {Mazzali}, {Iwamoto}, {Nomoto}, {Umeda}, {Nakamura}, {Young}, {Suzuki}, {Shigeyama}, {Koshut}, {Kippen}, {Robinson}, {de Wildt}, {Wijers}, {Tanvir}, {Greiner}, {Pian}, {Palazzi}, {Frontera}, {Masetti}, {Nicastro}, {Feroci}, {Costa}, {Piro}, {Peterson}, {Tinney}, {Boyle}, {Cannon}, {Stathakis}, {Sadler}, {Begam}, \& {Ianna}}]{1998Natur.395..670G}
{Galama}, T.~J., {Vreeswijk}, P.~M., {van Paradijs}, J., {et~al.} 1998, \nat, 395, 670, \dodoi{10.1038/27150}

\bibitem[{{Gehrels} \& {Razzaque}(2013)}]{2013FrPhy...8..661G}
{Gehrels}, N., \& {Razzaque}, S. 2013, Frontiers of Physics, 8, 661, \dodoi{10.1007/s11467-013-0282-3}

\bibitem[{{Gehrels} {et~al.}(2004){Gehrels}, {Chincarini}, {Giommi}, {Mason}, {Nousek}, {Wells}, {White}, {Barthelmy}, {Burrows}, {Cominsky}, {Hurley}, {Marshall}, {M{\'e}sz{\'a}ros}, {Roming}, {Angelini}, {Barbier}, {Belloni}, {Campana}, {Caraveo}, {Chester}, {Citterio}, {Cline}, {Cropper}, {Cummings}, {Dean}, {Feigelson}, {Fenimore}, {Frail}, {Fruchter}, {Garmire}, {Gendreau}, {Ghisellini}, {Greiner}, {Hill}, {Hunsberger}, {Krimm}, {Kulkarni}, {Kumar}, {Lebrun}, {Lloyd-Ronning}, {Markwardt}, {Mattson}, {Mushotzky}, {Norris}, {Osborne}, {Paczynski}, {Palmer}, {Park}, {Parsons}, {Paul}, {Rees}, {Reynolds}, {Rhoads}, {Sasseen}, {Schaefer}, {Short}, {Smale}, {Smith}, {Stella}, {Tagliaferri}, {Takahashi}, {Tashiro}, {Townsley}, {Tueller}, {Turner}, {Vietri}, {Voges}, {Ward}, {Willingale}, {Zerbi}, \& {Zhang}}]{2004ApJ...611.1005G}
{Gehrels}, N., {Chincarini}, G., {Giommi}, P., {et~al.} 2004, \apj, 611, 1005, \dodoi{10.1086/422091}

\bibitem[{{Golenetskii} {et~al.}(2015){Golenetskii}, {Aptekar}, {Frederiks}, {Pal'Shin}, {Oleynik}, {Ulanov}, {Svinkin}, {Tsvetkova}, {Lysenko}, {Kozlova}, \& {Cline}}]{2015GCN.18433....1G}
{Golenetskii}, S., {Aptekar}, R., {Frederiks}, D., {et~al.} 2015, GRB Coordinates Network, 18433, 1

\bibitem[{{Hamburg} {et~al.}(2019){Hamburg}, {Veres}, {Meegan}, {Burns}, {Connaughton}, {Goldstein}, {Kocevski}, \& {Roberts}}]{2019GCN.23707....1H}
{Hamburg}, R., {Veres}, P., {Meegan}, C., {et~al.} 2019, GRB Coordinates Network, 23707, 1

\bibitem[{{Hao} \& {Yuan}(2013)}]{2013ApJ...772...42H}
{Hao}, J.-M., \& {Yuan}, Y.-F. 2013, \apj, 772, 42, \dodoi{10.1088/0004-637X/772/1/42}

\bibitem[{{Hopkins} \& {Beacom}(2006)}]{2006ApJ...651..142H}
{Hopkins}, A.~M., \& {Beacom}, J.~F. 2006, \apj, 651, 142, \dodoi{10.1086/506610}

\bibitem[{{Hui}(2019)}]{2019GCN.24002....1H}
{Hui}, C.~M. 2019, GRB Coordinates Network, 24002, 1

\bibitem[{{Jia} {et~al.}(2022){Jia}, {Hu}, {Yang}, {Zhang}, \& {Wang}}]{2022MNRAS.516.2575J}
{Jia}, X.~D., {Hu}, J.~P., {Yang}, J., {Zhang}, B.~B., \& {Wang}, F.~Y. 2022, \mnras, 516, 2575, \dodoi{10.1093/mnras/stac2356}

\bibitem[{{Jin} {et~al.}(2020){Jin}, {Covino}, {Liao}, {Li}, {D'Avanzo}, {Fan}, \& {Wei}}]{2020NatAs...4...77J}
{Jin}, Z.-P., {Covino}, S., {Liao}, N.-H., {et~al.} 2020, Nature Astronomy, 4, 77, \dodoi{10.1038/s41550-019-0892-y}

\bibitem[{{Kaneko} {et~al.}(2006){Kaneko}, {Preece}, {Briggs}, {Paciesas}, {Meegan}, \& {Band}}]{2006ApJS..166..298K}
{Kaneko}, Y., {Preece}, R.~D., {Briggs}, M.~S., {et~al.} 2006, \apjs, 166, 298, \dodoi{10.1086/505911}

\bibitem[{{Kelly} {et~al.}(2008){Kelly}, {Kirshner}, \& {Pahre}}]{2008ApJ...687.1201K}
{Kelly}, P.~L., {Kirshner}, R.~P., \& {Pahre}, M. 2008, \apj, 687, 1201, \dodoi{10.1086/591925}

\bibitem[{{Kistler} {et~al.}(2008){Kistler}, {Y{\"u}ksel}, {Beacom}, \& {Stanek}}]{2008ApJ...673L.119K}
{Kistler}, M.~D., {Y{\"u}ksel}, H., {Beacom}, J.~F., \& {Stanek}, K.~Z. 2008, \apjl, 673, L119, \dodoi{10.1086/527671}

\bibitem[{{Kocevski} \& {Petrosian}(2013)}]{2013ApJ...765..116K}
{Kocevski}, D., \& {Petrosian}, V. 2013, \apj, 765, 116, \dodoi{10.1088/0004-637X/765/2/116}

\bibitem[{{Kouveliotou} {et~al.}(1993){Kouveliotou}, {Meegan}, {Fishman}, {Bhat}, {Briggs}, {Koshut}, {Paciesas}, \& {Pendleton}}]{1993ApJ...413L.101K}
{Kouveliotou}, C., {Meegan}, C.~A., {Fishman}, G.~J., {et~al.} 1993, \apjl, 413, L101, \dodoi{10.1086/186969}

\bibitem[{{Lan} {et~al.}(2022){Lan}, {Wei}, {Li}, {Zeng}, \& {Wu}}]{2022ApJ...938..129L}
{Lan}, G.-X., {Wei}, J.-J., {Li}, Y., {Zeng}, H.-D., \& {Wu}, X.-F. 2022, \apj, 938, 129, \dodoi{10.3847/1538-4357/ac8fec}

\bibitem[{{Lan} {et~al.}(2021){Lan}, {Wei}, {Zeng}, {Li}, \& {Wu}}]{2021MNRAS.508...52L}
{Lan}, G.-X., {Wei}, J.-J., {Zeng}, H.-D., {Li}, Y., \& {Wu}, X.-F. 2021, \mnras, 508, 52, \dodoi{10.1093/mnras/stab2508}

\bibitem[{{Lan} {et~al.}(2019){Lan}, {Zeng}, {Wei}, \& {Wu}}]{2019MNRAS.488.4607L}
{Lan}, G.-X., {Zeng}, H.-D., {Wei}, J.-J., \& {Wu}, X.-F. 2019, \mnras, 488, 4607, \dodoi{10.1093/mnras/stz2011}

\bibitem[{{Langer} {et~al.}(2010){Langer}, {van Marle}, \& {Yoon}}]{2010NewAR..54..206L}
{Langer}, N., {van Marle}, A.~J., \& {Yoon}, S.~C. 2010, \nar, 54, 206, \dodoi{10.1016/j.newar.2010.09.012}

\bibitem[{{Levan} {et~al.}(2024){Levan}, {Gompertz}, {Salafia}, {Bulla}, {Burns}, {Hotokezaka}, {Izzo}, {Lamb}, {Malesani}, {Oates}, {Ravasio}, {Rouco Escorial}, {Schneider}, {Sarin}, {Schulze}, {Tanvir}, {Ackley}, {Anderson}, {Brammer}, {Christensen}, {Dhillon}, {Evans}, {Fausnaugh}, {Fong}, {Fruchter}, {Fryer}, {Fynbo}, {Gaspari}, {Heintz}, {Hjorth}, {Kennea}, {Kennedy}, {Laskar}, {Leloudas}, {Mandel}, {Martin-Carrillo}, {Metzger}, {Nicholl}, {Nugent}, {Palmerio}, {Pugliese}, {Rastinejad}, {Rhodes}, {Rossi}, {Saccardi}, {Smartt}, {Stevance}, {Tohuvavohu}, {van der Horst}, {Vergani}, {Watson}, {Barclay}, {Bhirombhakdi}, {Breedt}, {Breeveld}, {Brown}, {Campana}, {Chrimes}, {D'Avanzo}, {D'Elia}, {De Pasquale}, {Dyer}, {Galloway}, {Garbutt}, {Green}, {Hartmann}, {Jakobsson}, {Kerry}, {Kouveliotou}, {Langeroodi}, {Le Floc'h}, {Leung}, {Littlefair}, {Munday}, {O'Brien}, {Parsons}, {Pelisoli}, {Sahman}, {Salvaterra}, {Sbarufatti}, {Steeghs}, {Tagliaferri}, {Th{\"o}ne}, {de Ugarte Postigo}, \&
  {Kann}}]{2024Natur.626..737L}
{Levan}, A.~J., {Gompertz}, B.~P., {Salafia}, O.~S., {et~al.} 2024, \nat, 626, 737, \dodoi{10.1038/s41586-023-06759-1}

\bibitem[{{Li} {et~al.}(2024){Li}, {Yang}, {Yi}, {Hu}, {Qu}, \& {Wang}}]{2024A&A...689A.165L}
{Li}, J.-L., {Yang}, Y.-P., {Yi}, S.-X., {et~al.} 2024, \aap, 689, A165, \dodoi{10.1051/0004-6361/202348542}

\bibitem[{{Li} {et~al.}(2023){Li}, {Yang}, {Yi}, {Hu}, {Wang}, \& {Qu}}]{2023ApJ...953...58L}
---. 2023, \apj, 953, 58, \dodoi{10.3847/1538-4357/ace107}

\bibitem[{{Li}(2008)}]{2008MNRAS.388.1487L}
{Li}, L.-X. 2008, \mnras, 388, 1487, \dodoi{10.1111/j.1365-2966.2008.13488.x}

\bibitem[{{Liddle}(2007)}]{2007MNRAS.377L..74L}
{Liddle}, A.~R. 2007, \mnras, 377, L74, \dodoi{10.1111/j.1745-3933.2007.00306.x}

\bibitem[{{Llamas Lanza} {et~al.}(2024){Llamas Lanza}, {Godet}, {Arcier}, {Yassine}, {Atteia}, \& {Bouchet}}]{2024arXiv240303266L}
{Llamas Lanza}, M., {Godet}, O., {Arcier}, B., {et~al.} 2024, arXiv e-prints, arXiv:2403.03266, \dodoi{10.48550/arXiv.2403.03266}

\bibitem[{{Lloyd-Ronning} {et~al.}(2019){Lloyd-Ronning}, {Aykutalp}, \& {Johnson}}]{2019MNRAS.488.5823L}
{Lloyd-Ronning}, N.~M., {Aykutalp}, A., \& {Johnson}, J.~L. 2019, \mnras, 488, 5823, \dodoi{10.1093/mnras/stz2155}

\bibitem[{{L{\"u}} {et~al.}(2010){L{\"u}}, {Liang}, {Zhang}, \& {Zhang}}]{2010ApJ...725.1965L}
{L{\"u}}, H.-J., {Liang}, E.-W., {Zhang}, B.-B., \& {Zhang}, B. 2010, \apj, 725, 1965, \dodoi{10.1088/0004-637X/725/2/1965}

\bibitem[{{Mailyan} \& {Goldstein}(2016)}]{2016GCN.20192....1M}
{Mailyan}, B., \& {Goldstein}, A. 2016, GRB Coordinates Network, 20192, 1

\bibitem[{{Markwardt} {et~al.}(2017){Markwardt}, {Barthelmy}, {Cummings}, {Gehrels}, {Krimm}, {Lien}, {Palmer}, {Sakamoto}, {Stamatikos}, \& {Ukwatta}}]{2017GCN.20456....1M}
{Markwardt}, C.~B., {Barthelmy}, S.~D., {Cummings}, J.~R., {et~al.} 2017, GRB Coordinates Network, 20456, 1

\bibitem[{{Marshall} {et~al.}(1983){Marshall}, {Tananbaum}, {Avni}, \& {Zamorani}}]{1983ApJ26935M}
{Marshall}, H.~L., {Tananbaum}, H., {Avni}, Y., \& {Zamorani}, G. 1983, \apj, 269, 35, \dodoi{10.1086/161016}

\bibitem[{{Mazets} {et~al.}(1981){Mazets}, {Golenetskii}, {Ilinskii}, {Panov}, {Aptekar}, {Gurian}, {Proskura}, {Sokolov}, {Sokolova}, \& {Kharitonova}}]{1981Ap&SS..80....3M}
{Mazets}, E.~P., {Golenetskii}, S.~V., {Ilinskii}, V.~N., {et~al.} 1981, \apss, 80, 3, \dodoi{10.1007/BF00649140}

\bibitem[{{Minaev} \& {Pozanenko}(2020)}]{2020MNRAS.492.1919M}
{Minaev}, P.~Y., \& {Pozanenko}, A.~S. 2020, \mnras, 492, 1919, \dodoi{10.1093/mnras/stz3611}

\bibitem[{{Moss} {et~al.}(2022){Moss}, {Lien}, {Guiriec}, {Cenko}, \& {Sakamoto}}]{2022ApJ...927..157M}
{Moss}, M., {Lien}, A., {Guiriec}, S., {Cenko}, S.~B., \& {Sakamoto}, T. 2022, \apj, 927, 157, \dodoi{10.3847/1538-4357/ac4d94}

\bibitem[{{Narumoto} \& {Totani}(2006)}]{2006ApJ...643...81N}
{Narumoto}, T., \& {Totani}, T. 2006, \apj, 643, 81, \dodoi{10.1086/502708}

\bibitem[{{Nava} {et~al.}(2012){Nava}, {Salvaterra}, {Ghirlanda}, {Ghisellini}, {Campana}, {Covino}, {Cusumano}, {D'Avanzo}, {D'Elia}, {Fugazza}, {Melandri}, {Sbarufatti}, {Vergani}, \& {Tagliaferri}}]{2012MNRAS.421.1256N}
{Nava}, L., {Salvaterra}, R., {Ghirlanda}, G., {et~al.} 2012, \mnras, 421, 1256, \dodoi{10.1111/j.1365-2966.2011.20394.x}

\bibitem[{{Palmer} {et~al.}(2018){Palmer}, {Barthelmy}, {Cummings}, {Krimm}, {Lien}, {Markwardt}, {Racusin}, {Sakamoto}, {Stamatikos}, \& {Ukwatta}}]{2018GCN.22566....1P}
{Palmer}, D.~M., {Barthelmy}, S.~D., {Cummings}, J.~R., {et~al.} 2018, GRB Coordinates Network, 22566, 1

\bibitem[{{Pescalli} {et~al.}(2015){Pescalli}, {Ghirlanda}, {Salafia}, {Ghisellini}, {Nappo}, \& {Salvaterra}}]{2015MNRAS.447.1911P}
{Pescalli}, A., {Ghirlanda}, G., {Salafia}, O.~S., {et~al.} 2015, \mnras, 447, 1911, \dodoi{10.1093/mnras/stu2482}

\bibitem[{{Pescalli} {et~al.}(2016){Pescalli}, {Ghirlanda}, {Salvaterra}, {Ghisellini}, {Vergani}, {Nappo}, {Salafia}, {Melandri}, {Covino}, \& {G{\"o}tz}}]{2016A&A...587A..40P}
{Pescalli}, A., {Ghirlanda}, G., {Salvaterra}, R., {et~al.} 2016, \aap, 587, A40, \dodoi{10.1051/0004-6361/201526760}

\bibitem[{{Petrosian} \& {Dainotti}(2024)}]{2024ApJ...963L..12P}
{Petrosian}, V., \& {Dainotti}, M.~G. 2024, \apjl, 963, L12, \dodoi{10.3847/2041-8213/ad2763}

\bibitem[{{Petrosian} {et~al.}(2015){Petrosian}, {Kitanidis}, \& {Kocevski}}]{2015ApJ...806...44P}
{Petrosian}, V., {Kitanidis}, E., \& {Kocevski}, D. 2015, \apj, 806, 44, \dodoi{10.1088/0004-637X/806/1/44}

\bibitem[{{Poolakkil} {et~al.}(2019{\natexlab{a}}){Poolakkil}, {Meegan}, \& {Fermi GBM Team}}]{2019GCN.24816....1P}
{Poolakkil}, S., {Meegan}, C., \& {Fermi GBM Team}. 2019{\natexlab{a}}, GRB Coordinates Network, 24816, 1

\bibitem[{{Poolakkil} {et~al.}(2019{\natexlab{b}}){Poolakkil}, {Meegan}, \& {Fermi GBM Team}}]{2019GCN.25130....1P}
---. 2019{\natexlab{b}}, GRB Coordinates Network, 25130, 1

\bibitem[{{Qu} {et~al.}(2019){Qu}, {Zeng}, \& {Yan}}]{2019MNRAS.490..758Q}
{Qu}, Y., {Zeng}, H., \& {Yan}, D. 2019, \mnras, 490, 758, \dodoi{10.1093/mnras/stz2651}

\bibitem[{{Qu} {et~al.}(2024){Qu}, {Man}, {Yi}, \& {Yang}}]{2024ApJ...976..170Q}
{Qu}, Y.-K., {Man}, Z.-X., {Yi}, S.-X., \& {Yang}, Y.-P. 2024, \apj, 976, 170, \dodoi{10.3847/1538-4357/ad88e7}

\bibitem[{{Riess} {et~al.}(1998){Riess}, {Filippenko}, {Challis}, {Clocchiatti}, {Diercks}, {Garnavich}, {Gilliland}, {Hogan}, {Jha}, {Kirshner}, {Leibundgut}, {Phillips}, {Reiss}, {Schmidt}, {Schommer}, {Smith}, {Spyromilio}, {Stubbs}, {Suntzeff}, \& {Tonry}}]{1998AJ....116.1009R}
{Riess}, A.~G., {Filippenko}, A.~V., {Challis}, P., {et~al.} 1998, \aj, 116, 1009, \dodoi{10.1086/300499}

\bibitem[{{Sakamoto} {et~al.}(2011){Sakamoto}, {Barthelmy}, {Baumgartner}, {Cummings}, {Fenimore}, {Gehrels}, {Krimm}, {Markwardt}, {Palmer}, {Parsons}, {Sato}, {Stamatikos}, {Tueller}, {Ukwatta}, \& {Zhang}}]{2011ApJS..195....2S}
{Sakamoto}, T., {Barthelmy}, S.~D., {Baumgartner}, W.~H., {et~al.} 2011, \apjs, 195, 2, \dodoi{10.1088/0067-0049/195/1/2}

\bibitem[{{Salvaterra} {et~al.}(2009){Salvaterra}, {Della Valle}, {Campana}, {Chincarini}, {Covino}, {D'Avanzo}, {Fern{\'a}ndez-Soto}, {Guidorzi}, {Mannucci}, {Margutti}, {Th{\"o}ne}, {Antonelli}, {Barthelmy}, {de Pasquale}, {D'Elia}, {Fiore}, {Fugazza}, {Hunt}, {Maiorano}, {Marinoni}, {Marshall}, {Molinari}, {Nousek}, {Pian}, {Racusin}, {Stella}, {Amati}, {Andreuzzi}, {Cusumano}, {Fenimore}, {Ferrero}, {Giommi}, {Guetta}, {Holland}, {Hurley}, {Israel}, {Mao}, {Markwardt}, {Masetti}, {Pagani}, {Palazzi}, {Palmer}, {Piranomonte}, {Tagliaferri}, \& {Testa}}]{2009Natur.461.1258S}
{Salvaterra}, R., {Della Valle}, M., {Campana}, S., {et~al.} 2009, \nat, 461, 1258, \dodoi{10.1038/nature08445}

\bibitem[{{Salvaterra} {et~al.}(2012){Salvaterra}, {Campana}, {Vergani}, {Covino}, {D'Avanzo}, {Fugazza}, {Ghirlanda}, {Ghisellini}, {Melandri}, {Nava}, {Sbarufatti}, {Flores}, {Piranomonte}, \& {Tagliaferri}}]{2012ApJ...749...68S}
{Salvaterra}, R., {Campana}, S., {Vergani}, S.~D., {et~al.} 2012, \apj, 749, 68, \dodoi{10.1088/0004-637X/749/1/68}

\bibitem[{{Sun} {et~al.}(2015){Sun}, {Zhang}, \& {Li}}]{2015ApJ...812...33S}
{Sun}, H., {Zhang}, B., \& {Li}, Z. 2015, \apj, 812, 33, \dodoi{10.1088/0004-637X/812/1/33}

\bibitem[{{Svensson} {et~al.}(2010){Svensson}, {Levan}, {Tanvir}, {Fruchter}, \& {Strolger}}]{2010MNRAS.405...57S}
{Svensson}, K.~M., {Levan}, A.~J., {Tanvir}, N.~R., {Fruchter}, A.~S., \& {Strolger}, L.~G. 2010, \mnras, 405, 57, \dodoi{10.1111/j.1365-2966.2010.16442.x}

\bibitem[{{Svinkin} {et~al.}(2019){Svinkin}, {Golenetskii}, {Aptekar}, {Frederiks}, {Ulanov}, {Tsvetkova}, {Lysenko}, {Kozlova}, {Cline}, \& {Konus-Wind Team}}]{2019GCN.25974....1S}
{Svinkin}, D., {Golenetskii}, S., {Aptekar}, R., {et~al.} 2019, GRB Coordinates Network, 25974, 1

\bibitem[{{Tanvir} {et~al.}(2013){Tanvir}, {Levan}, {Fruchter}, {Hjorth}, {Hounsell}, {Wiersema}, \& {Tunnicliffe}}]{2013Natur.500..547T}
{Tanvir}, N.~R., {Levan}, A.~J., {Fruchter}, A.~S., {et~al.} 2013, \nat, 500, 547, \dodoi{10.1038/nature12505}

\bibitem[{{Tian} {et~al.}(2023){Tian}, {Li}, {Yi}, {Yang}, {Hu}, {Qu}, \& {Wang}}]{2023ApJ...958...74T}
{Tian}, X., {Li}, J.-L., {Yi}, S.-X., {et~al.} 2023, \apj, 958, 74, \dodoi{10.3847/1538-4357/acfed8}

\bibitem[{{Totani}(1997)}]{1997ApJ...486L..71T}
{Totani}, T. 1997, \apjl, 486, L71, \dodoi{10.1086/310853}

\bibitem[{{Troja} {et~al.}(2019){Troja}, {Castro-Tirado}, {Becerra Gonz{\'a}lez}, {Hu}, {Ryan}, {Cenko}, {Ricci}, {Novara}, {S{\'a}nchez-R{\'a}mirez}, {Acosta-Pulido}, {Ackley}, {Caballero Garc{\'\i}a}, {Eikenberry}, {Guziy}, {Jeong}, {Lien}, {M{\'a}rquez}, {Pandey}, {Park}, {Sakamoto}, {Tello}, {Sokolov}, {Sokolov}, {Tiengo}, {Valeev}, {Zhang}, \& {Veilleux}}]{2019MNRAS.489.2104T}
{Troja}, E., {Castro-Tirado}, A.~J., {Becerra Gonz{\'a}lez}, J., {et~al.} 2019, \mnras, 489, 2104, \dodoi{10.1093/mnras/stz2255}

\bibitem[{{Tsvetkova} {et~al.}(2017){Tsvetkova}, {Frederiks}, {Golenetskii}, {Lysenko}, {Oleynik}, {Pal'shin}, {Svinkin}, {Ulanov}, {Cline}, {Hurley}, \& {Aptekar}}]{2017ApJ...850..161T}
{Tsvetkova}, A., {Frederiks}, D., {Golenetskii}, S., {et~al.} 2017, \apj, 850, 161, \dodoi{10.3847/1538-4357/aa96af}

\bibitem[{{Tsvetkova} {et~al.}(2018{\natexlab{a}}){Tsvetkova}, {Golenetskii}, {Aptekar}, {Frederiks}, {Oleynik}, {Ulanov}, {Svinkin}, {Lysenko}, {Kozlova}, \& {Cline}}]{2018GCN.22513....1T}
{Tsvetkova}, A., {Golenetskii}, S., {Aptekar}, R., {et~al.} 2018{\natexlab{a}}, GRB Coordinates Network, 22513, 1

\bibitem[{{Tsvetkova} {et~al.}(2018{\natexlab{b}}){Tsvetkova}, {Golenetskii}, {Aptekar}, {Frederiks}, {Ulanov}, {Svinkin}, {Lysenko}, {Kozlova}, \& {Cline}}]{2018GCN.23363....1T}
---. 2018{\natexlab{b}}, GRB Coordinates Network, 23363, 1

\bibitem[{{Tsvetkova} {et~al.}(2019){Tsvetkova}, {Golenetskii}, {Aptekar}, {Frederiks}, {Ulanov}, {Svinkin}, {Lysenko}, {Kozlova}, \& {Cline}}]{2019GCN.23637....1T}
---. 2019, GRB Coordinates Network, 23637, 1

\bibitem[{{Veres} \& {Mailyan}(2018)}]{2018GCN.23053....1P}
{Veres}, P.~{Meegan}, C., \& {Mailyan}, B. 2018, GRB Coordinates Network, 23053, 1

\bibitem[{{Virgili} {et~al.}(2011){Virgili}, {Zhang}, {Nagamine}, \& {Choi}}]{2011MNRAS.417.3025V}
{Virgili}, F.~J., {Zhang}, B., {Nagamine}, K., \& {Choi}, J.-H. 2011, \mnras, 417, 3025, \dodoi{10.1111/j.1365-2966.2011.19459.x}

\bibitem[{{von Kienlin}(2018)}]{2018GCN.22386....1V}
{von Kienlin}, A. 2018, GRB Coordinates Network, 22386, 1

\bibitem[{{von Kienlin} \& {Burns}(2015)}]{2015GCN.17319....1V}
{von Kienlin}, A., \& {Burns}, E. 2015, GRB Coordinates Network, 17319, 1

\bibitem[{{von Kienlin} {et~al.}(2020){von Kienlin}, {Meegan}, {Paciesas}, {Bhat}, {Bissaldi}, {Briggs}, {Burns}, {Cleveland}, {Gibby}, {Giles}, {Goldstein}, {Hamburg}, {Hui}, {Kocevski}, {Mailyan}, {Malacaria}, {Poolakkil}, {Preece}, {Roberts}, {Veres}, \& {Wilson-Hodge}}]{2020ApJ...893...46V}
{von Kienlin}, A., {Meegan}, C.~A., {Paciesas}, W.~S., {et~al.} 2020, \apj, 893, 46, \dodoi{10.3847/1538-4357/ab7a18}

\bibitem[{{Wang} {et~al.}(2024){Wang}, {Tan}, {Xiong}, {Yi}, {Moradi}, {Li}, {Zhang}, {Wang}, {Meng}, {Liu}, {Wang}, {Xie}, {Xue}, {Yu}, {Zhang}, {Zhang}, {Zhang}, \& {Zheng}}]{2024arXiv240702376W}
{Wang}, C.-W., {Tan}, W.-J., {Xiong}, S.-L., {et~al.} 2024, arXiv e-prints, arXiv:2407.02376, \dodoi{10.48550/arXiv.2407.02376}

\bibitem[{{Wang} {et~al.}(2020){Wang}, {Zou}, {Liu}, {Liao}, {Liu}, {Chai}, \& {Xia}}]{2020ApJ...893...77W}
{Wang}, F., {Zou}, Y.-C., {Liu}, F., {et~al.} 2020, \apj, 893, 77, \dodoi{10.3847/1538-4357/ab0a86}

\bibitem[{{Wang} \& {Dai}(2009)}]{2009MNRAS.400L..10W}
{Wang}, F.~Y., \& {Dai}, Z.~G. 2009, \mnras, 400, L10, \dodoi{10.1111/j.1745-3933.2009.00751.x}

\bibitem[{{Wang} {et~al.}(2015){Wang}, {Dai}, \& {Liang}}]{2015NewAR..67....1W}
{Wang}, F.~Y., {Dai}, Z.~G., \& {Liang}, E.~W. 2015, \nar, 67, 1, \dodoi{10.1016/j.newar.2015.03.001}

\bibitem[{{Wang} {et~al.}(2022){Wang}, {Hu}, {Zhang}, \& {Dai}}]{2022ApJ...924...97W}
{Wang}, F.~Y., {Hu}, J.~P., {Zhang}, G.~Q., \& {Dai}, Z.~G. 2022, \apj, 924, 97, \dodoi{10.3847/1538-4357/ac3755}

\bibitem[{{Wang} {et~al.}(2016){Wang}, {Wang}, {Cheng}, \& {Dai}}]{2016A&A...585A..68W}
{Wang}, J.~S., {Wang}, F.~Y., {Cheng}, K.~S., \& {Dai}, Z.~G. 2016, \aap, 585, A68, \dodoi{10.1051/0004-6361/201526485}

\bibitem[{{Wei} {et~al.}(2016){Wei}, {Cordier}, {Antier}, {Antilogus}, {Atteia}, {Bajat}, {Basa}, {Beckmann}, {Bernardini}, {Boissier}, {Bouchet}, {Burwitz}, {Claret}, {Dai}, {Daigne}, {Deng}, {Dornic}, {Feng}, {Foglizzo}, {Gao}, {Gehrels}, {Godet}, {Goldwurm}, {Gonzalez}, {Gosset}, {G{\"o}tz}, {Gouiffes}, {Grise}, {Gros}, {Guilet}, {Han}, {Huang}, {Huang}, {Jouret}, {Klotz}, {La Marle}, {Lachaud}, {Le Floch}, {Lee}, {Leroy}, {Li}, {Li}, {Li}, {Liang}, {Lyu}, {Mercier}, {Migliori}, {Mochkovitch}, {O'Brien}, {Osborne}, {Paul}, {Perinati}, {Petitjean}, {Piron}, {Qiu}, {Rau}, {Rodriguez}, {Schanne}, {Tanvir}, {Vangioni}, {Vergani}, {Wang}, {Wang}, {Wang}, {Wang}, {Watson}, {Webb}, {Wei}, {Willingale}, {Wu}, {Wu}, {Xin}, {Xu}, {Yu}, {Yu}, {Yu}, {Zhang}, {Zhang}, {Zhang}, \& {Zhou}}]{2016arXiv161006892W}
{Wei}, J., {Cordier}, B., {Antier}, S., {et~al.} 2016, arXiv e-prints, arXiv:1610.06892, \dodoi{10.48550/arXiv.1610.06892}

\bibitem[{{Yang} {et~al.}(2022){Yang}, {Ai}, {Zhang}, {Zhang}, {Liu}, {Wang}, {Yang}, {Yin}, {Li}, \& {L{\"u}}}]{2022Natur.612..232Y}
{Yang}, J., {Ai}, S., {Zhang}, B.-B., {et~al.} 2022, \nat, 612, 232, \dodoi{10.1038/s41586-022-05403-8}

\bibitem[{{Yonetoku} {et~al.}(2004){Yonetoku}, {Murakami}, {Nakamura}, {Yamazaki}, {Inoue}, \& {Ioka}}]{2004ApJ...609..935Y}
{Yonetoku}, D., {Murakami}, T., {Nakamura}, T., {et~al.} 2004, \apj, 609, 935, \dodoi{10.1086/421285}

\bibitem[{{Yu} {et~al.}(2015){Yu}, {Wang}, {Dai}, \& {Cheng}}]{2015ApJS..218...13Y}
{Yu}, H., {Wang}, F.~Y., {Dai}, Z.~G., \& {Cheng}, K.~S. 2015, \apjs, 218, 13, \dodoi{10.1088/0067-0049/218/1/13}

\bibitem[{{Yu}(2014)}]{2014GCN.17216....1Y}
{Yu}, H.~F. 2014, GRB Coordinates Network, 17216, 1

\bibitem[{{Y{\"u}ksel} {et~al.}(2008){Y{\"u}ksel}, {Kistler}, {Beacom}, \& {Hopkins}}]{2008ApJ...683L...5Y}
{Y{\"u}ksel}, H., {Kistler}, M.~D., {Beacom}, J.~F., \& {Hopkins}, A.~M. 2008, \apjl, 683, L5, \dodoi{10.1086/591449}

\bibitem[{{Zeng} {et~al.}(2016){Zeng}, {Melia}, \& {Zhang}}]{2016MNRAS.462.3094Z}
{Zeng}, H., {Melia}, F., \& {Zhang}, L. 2016, \mnras, 462, 3094, \dodoi{10.1093/mnras/stw1817}

\bibitem[{{Zeng} {et~al.}(2014){Zeng}, {Yan}, \& {Zhang}}]{2014MNRAS.441.1760Z}
{Zeng}, H., {Yan}, D., \& {Zhang}, L. 2014, \mnras, 441, 1760, \dodoi{10.1093/mnras/stu644}

\bibitem[{{Zhang}(2025)}]{2025JHEAp..45..325Z}
{Zhang}, B. 2025, Journal of High Energy Astrophysics, 45, 325, \dodoi{10.1016/j.jheap.2024.12.013}

\bibitem[{{Zhang} {et~al.}(2009){Zhang}, {Zhang}, {Virgili}, {Liang}, {Kann}, {Wu}, {Proga}, {Lv}, {Toma}, {M{\'e}sz{\'a}ros}, {Burrows}, {Roming}, \& {Gehrels}}]{2009ApJ...703.1696Z}
{Zhang}, B., {Zhang}, B.-B., {Virgili}, F.~J., {et~al.} 2009, \apj, 703, 1696, \dodoi{10.1088/0004-637X/703/2/1696}

\bibitem[{{Zhang}(2014)}]{2014GCN.16798....1Z}
{Zhang}, B.~B. 2014, GRB Coordinates Network, 16798, 1

\bibitem[{{Zhang} \& {Wang}(2018)}]{Zhang2018}
{Zhang}, G.~Q., \& {Wang}, F.~Y. 2018, \apj, 852, 1, \dodoi{10.3847/1538-4357/aa9ce5}

\end{thebibliography}
\bibliographystyle{aasjournal}



\end{document}